\documentclass[12pt,preprint]{emulateapj}
\begin{document}
\title{Very Low-luminosity galaxies in the early universe have
  observed sizes similar to single star cluster complexes}

\author{R.J. Bouwens\altaffilmark{1}, G.D. Illingworth\altaffilmark{3}, P.A. Oesch\altaffilmark{4}, M. Maseda\altaffilmark{1}, B. Ribeiro\altaffilmark{1,5}, M. Stefanon\altaffilmark{1,2}, D. Lam\altaffilmark{1}}
\altaffiltext{1}{Leiden Observatory,
  Leiden University, NL-2300 RA Leiden, Netherlands}
\altaffiltext{2}{Department of Astronomy, Yale University, New Haven,
  CT 06520}
\altaffiltext{3}{UCO/Lick Observatory, University of California, Santa
  Cruz, CA 95064}
\altaffiltext{4}{Observatoire de Gen{\`e}ve, 1290 Versoix, Switzerland}
\altaffiltext{5}{Aix Marseille Univ, CNRS, LAM, Laboratoire d’Astrophysique de Marseille, Marseille, France}
\begin{abstract}
We compare the sizes and luminosities of 307 faint z = 6-8 sources
revealed by the Hubble Frontier Fields (HFF) program with sources in
the nearby universe. Making use of the latest lensing models and data
from the first four HFF clusters with an extensive suite of public
lens models, we measure both the sizes and luminosities for 153
$z\sim6$, 101 $z\sim7$, and 53 $z\sim8$ galaxies.  The sizes range
over more than a decade from $\sim$500 to $<$50 pc.  Extremely small
sizes are inferred for many of our lowest luminosity sources, reaching
individual sizes as small as 10-30 pc (the smallest is 11$_{-6}^{+28}$
pc). The uncertainty in these measures ranges from 80 pc for the
largest sources to typically about 20 pc for the smallest.  Such sizes
are smaller than extrapolations of the size-luminosity relation, and
expectations for the completeness of our faint samples, suggesting a
likely break in the size-luminosity relation at $\sim$$-$17 mag with
$r \propto L^{0.50_{-0.11}^{+0.10}}$.  The sizes and luminosities of the
lowest-luminosity sources are similar to those of single star cluster
complexes like 30 Doradus in the lower-redshift universe and -- in a
few cases -- super star clusters.  Remarkably, our identification of
these compact, faint star-forming sources in the $z\sim6$-8 universe
also allow us to set upper limits on the proto-globular cluster LF at
$z\sim6$.  Comparisons with recent models allow us to rule out (with
some caveats) some scenarios for proto-globular cluster formation and
set useful upper limits on other less extreme ones.  Our results
suggest we may be very close to discovering a bona-fide population of
forming globular clusters at high redshift.
\end{abstract}

\section{Introduction}

There are a wide variety of evolved stellar systems in the nearby
universe (Norris et al.\ 2014), from globular clusters (Brodie \&
Strader 2006; Kruijssen 2014; Renzini et al.\ 2015) to compact
elliptical galaxies (e.g., Faber 1973) to ultra-faint dwarfs (e.g.,
Simon \& Geha 2007) to ultra-diffuse spheroids (e.g., van Dokkum et
al.\ 2015), each of which presumably has its own characteristic
formation pathway.  The high stellar densities in many of these
systems in combination with their old ages (e.g., Forbes \& Bridges
2010) suggest that the majority of their star formation occured at
$z\gtrsim 1.5$ when the gas densities in the universe were in general
much higher.

One potentially promising way forward to investigate the formation of
these local systems is by obtaining a sensitive, high-resolution view
into the distant universe.  Fortunately, such observations can be
obtained by combining the power of long exposures with the Hubble
Space Telescope with the magnifying effect of gravitational lensing,
as recently implemented in the ambitious Hubble Frontier Fields (HFF)
program (Coe et al.\ 2015; Lotz et al.\ 2017).  Indeed the HFF program
has great potential to examine the structure and morphology of faint
high-redshift galaxies in great detail.  Sources can be stretched by
factors of 5 to 20 along one of their axes, allowing the structure in
such systems to be studied at very high spatial resolution.  One
significant earlier example of what could be done was the
highly-magnified $z=4.92$ galaxy behind MS1358+62 (Franx et al.\ 1997;
Swinbank et al.\ 2009) where star-forming clumps just 200 pc in size
could be partially resolved.

Already there have been several uses of the HFF observations to look
in detail at the size distribution of extremely faint galaxies.  In an
early study leveraging HFF observations over the first HFF cluster
Abell 2744, Kawamata et al.\ (2015) made use of the data to map out
the distribution of galaxy sizes vs. luminosities, while Laporte et
al.\ (2016) looked further into the sizes of fainter galaxies using
the HFF data over the second and third HFF clusters.  Interestingly
enough, Kawamata et al.\ (2015) identified a few $\sim-17$ mag
sources\footnote{Specifically HFF1C-i10 and HFF1C-i13 from Kawamata et
  al.\ (2015).} with nominal physical sizes less than 40 pc using
their own lensing model (Ishigaki et al.\ 2015).

In Bouwens et al.\ (2017a), we pursued constraints on the physical
sizes of fainter $>$$-16.5$ mag $z=2$-8 galaxies in the HFF
observations, looking at both (1) the prevalence of sources as a
function of lensing shear and (2) detailed size constraints on sources
in particularly high magnification areas.  These analyses provided the
first evidence that very low luminosity ($>$$-16.5$ mag) galaxies
might have especially small sizes, i.e., in the range of tens of
parsecs to over 100 pc.  This is very similar to the sizes of
molecular clouds and star cluster complexes in the $z\sim0$-3 universe
(e.g., Kennicutt et al.\ 2003; Bastian et al.\ 2006; Jones et
al.\ 2010; Swinbank et al.\ 2012; Adamo et al.\ 2013; Johnson et
al.\ 2017; Dessauges-Zavadsky et al.\ 2017).

Finally, Vanzella et al.\ (2017a) made use of the HFF observations and
their own lensing magnification models (Caminha et al.\ 2016) to
identify a set of very small sources in the $z\sim3$-6 universe, which
Vanzella et al.\ (2017a) speculated could correspond to proto-globular
clusters.  To support such a characterization, Vanzella et
al.\ (2017a) made use of the available MUSE spectroscopy on the
sources, noting small probable dynamical masses (due to the small
measured velocity dispersions) and probable physical associations with
brighter neighbors (due to their similar redshifts).  In a follow-up
analysis, Vanzella et al.\ (2017b) identified two candidate super star
clusters at $z=3.222$ with sizes of 30$\pm$11 pc they inferred to be
associated with a brighter neighboring galaxy.

The purpose of the present work is to take one step beyond these
studies, using large samples of $z=6$-8 galaxies to map out the size
distribution to very low luminosities,\footnote{See also new work by
  Kawamata et al.\ (2017).} and exploring the connection with stellar
systems at lower redshifts.  In doing so, we make use of the first
four clusters from the HFF program which have the most refined set of
gravitational lensing models, select $z=6$-8 galaxies behind them, and
then measure sizes for individual lensed galaxies.  Throughout the
paper, we assume a standard ``concordance'' cosmology with $H_0=70$ km
s$^{-1}$ Mpc$^{-1}$, $\Omega_{\rm m}=0.3$ and $\Omega_{\Lambda}=0.7$,
which is in good agreement with recent cosmological constraints
(Planck Collaboration et al.\ 2015).  Magnitudes are in the AB system
(Oke \& Gunn 1983).

\section{Data Sets and Samples}

In our analysis, we make use of the v1.0 reductions of the HST
observations over the first four clusters that make up the Hubble
Frontier Fields program (Coe et al.\ 2015; Lotz et al.\ 2017).  These
reductions include all 140 orbits of HST imaging observations obtained
over each cluster (70 optical/ACS, 70 near-IR/WFC3/IR) plus any
additional archival observations taken over each cluster as a result
of other programs, e.g., CLASH (Postman et al.\ 2012) or GLASS
(Schmidt et al.\ 2014).  We focus on results from the first four
clusters because version 3 and version 4 public magnification models
are already available for those clusters, including multiple image
systems identified using the full HFF data set and substantial
spectroscopic redshift constraints on multiple image systems (Mahler
et al.\ 2017; Caminha et al.\ 2017; Schmidt et al.\ 2014; Vanzella et
al.\ 2014; Limousin et al.\ 2016; Jauzac et al.\ 2016; Owers et
al.\ 2011).

\begin{figure*}
\epsscale{1.11}
\plotone{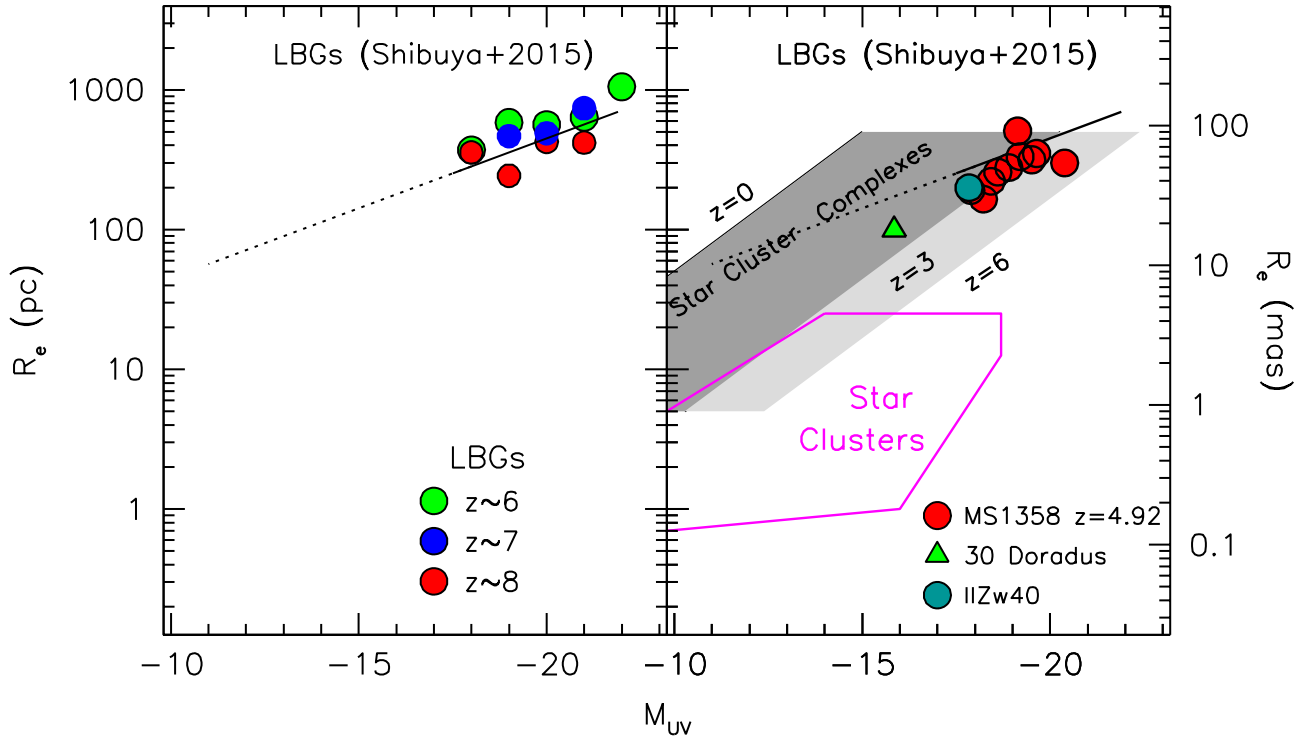}
\caption{Median size vs. luminosity relation of galaxies identified in
  blank field studies, i.e., the HUDF and CANDELS (\textit{left
    panel}) and those of star cluster complexes (\textit{right
    panel}).  The canonical size-luminosity relation is presented
  using both the Shibuya et al.\ (2015) fit results (\textit{black
    line}) and median sizes at $z\sim6$ (\textit{green circles}),
  $z\sim7$ (\textit{blue circles}), and $z\sim8$ (\textit{red
    circles}).  The black dotted line shows an extrapolation of the
  best-fit Shibuya et al.\ (2015) trend to lower luminosities.  The
  dark gray region indicates the size-luminosity relation for star
  cluster complexes in $z=0$-3 galaxies inferred by Livermore et
  al.\ (2015) by fitting to the $z=0$ results from SINGS (Kennicutt et
  al.\ 2003), as well as the results of Jones et al.\ (2010), Swinbank
  et al.\ (2012), Livermore et al.\ (2012), Wisnioski et al.\ (2012),
  and Livermore et al.\ (2015).  The light gray region indicates the
  size-luminosity relation for star cluster complexes extrapolating
  this relation to $z=3$-6.  The solid red circles correspond to the
  measured sizes and equivalent $UV$ luminosities of the star cluster
  complexes identified in the highly magnified $z=4.92$ galaxy behind
  MS1358+62 (Franx et al.\ 1997) by Swinbank et al.\ (2009) and Jones
  et al.\ (2010), while the solid green triangle and cyan circle
  correspond to the sizes and luminosities of 30 Doradus and IIZw40,
  respectively (English \& Freeman 2003; Vanzi et al.\ 2008).  The
  magenta lines enclose the luminosities and sizes measured for star
  clusters and super star clusters at $z\sim0$ (Meurer et
  al.\ 1995).\label{fig:msre00}}
\end{figure*}

Before constructing catalogs of sources behind these clusters, the
subtraction of intracluster light and light from the brightest galaxy
was performed using \textsc{galfit} (Peng et al.\ 2002) and a
median-smoothing algorithm, which will be described in detail in
R.J. Bouwens et al.\ (2017, in prep).  As discussed in Appendix A of
Bouwens et al.\ (2017b), our procedure works at least as well as
similar procedures in Merlin et al.\ (2016) and Livermore et
al.\ (2017).

After modeling and subtracting light from the foreground cluster and
galaxies from the images, we move onto the selection of faint
high-redshift sources.  Here we restrict our focus to the selection of
sources at $z\sim6$, $z\sim7$, and $z\sim8$ because of the large
number of sources in those samples and because they can be selected
more securely than sources at lower redshifts.  Some further testing
we have performed on the faintest sources (i.e., $H_{160,AB}>28$) in
$z\sim5$ selections over the HFFs suggest they can be subject to
modest contamination from foreground cluster galaxies near the cluster
centers.  This could be worrisome, since that is where the model
magnification factors are frequently high.

The selection of sources in our $z\sim6$, $z\sim7$, and $z\sim8$
samples will be described in R.J. Bouwens et al.\ (2017, in prep) and
rely on the use of two color criteria and optical non-detection
criteria, as well as the integrated high and low redshift probability
computed with the photometric redshift code EAZY (Brammer et
al.\ 2008).  Our procedure for selecting $z\sim6$ galaxies is almost
identical to that already described in Bouwens et al.\ (2017b), while
our procedure for selecting $z\sim7$ and $z\sim8$ galaxies is similar
to that described in Bouwens et al.\ (2015).  Our $z\sim6$, $z\sim7$,
and $z\sim8$ samples from the first four HFF clusters contain 153,
101, and 53 sources, respectively, for a total of 307.

\section{Reference Size-Luminosity Relations}

To provide context for the measurements we obtain of the size and
luminosities of faint $z=6$-8 in the HFF observations (\S4), we first
provide a brief summary of the general size constraints that exist for
galaxies from field studies (\S3.1) while reviewing the size
measurements that have been made for star cluster complexes in the
redshift range $z\sim0$-3 (\S3.2) and star clusters at $z\sim0$
(\S3.3).

\subsection{Size-Luminosity Relation for Star Forming Galaxies at $z$$\sim$6-8 from Blank Field Studies}

It is useful for us to frame the constraints we obtain here for lensed
sources in our fields relative to the sizes of galaxies identified in
an extensive set of blank field studies (e.g., Ferguson et al.\ 2004;
Bouwens et al.\ 2004; Oesch et al.\ 2010; Grazian et al.\ 2012; Huang
et al.\ 2013; Ono et al.\ 2013; Shibuya et al.\ 2015; Holwerda et
al.\ 2015).

The most recent and comprehensive of these determinations is by
Shibuya et al.\ (2015), who conduct size measurements on $\sim$190,000
$z=0$-10 galaxies identified over the HUDF, the HUDF parallel fields,
the 5 CANDELS fields, and two of the HFF parallel fields.  The median
half-light radius of sources that Shibuya et al.\ (2015) measure for
their $z\sim6$ and $z\sim7$ samples is presented in the left panel of
Figure~\ref{fig:msre00} with blue and green circles, respectively, and
is well represented by the following relationship:
\begin{equation}
\log_{10} (r_e/\textrm{pc}) = (-0.4)(0.25)(M_{UV}+21) + 2.74
\end{equation}
where $r_e$ is the half-light radius in pc and $M_{UV}$ is the $UV$
luminosity at $\sim$1600\AA.  The above size-luminosity is included in
Figure~\ref{fig:msre00} as a solid black line over the range where
current observations provide a direct constraint on the relationship
and extrapolated to lower luminosities assuming the same slope
(\textit{dotted line}).

The Shibuya et al.\ (2015) size-luminosity relation is fairly typical
that seen in other studies (Mosleh et al.\ 2012; Huang et al.\ 2013;
van der Wel et al.\ 2014) for luminous galaxies across a range of
redshifts, from $z\sim2$ to $z\sim6$.  

It is valuable to recognize that current blank-field observations only
probe the high end of the luminosity range examined in this study.

\subsection{Size-Luminosity Relations for Star Cluster Complexes at $z<3$}

A second valuable reference point for the size measurements we will
make for faint $z=6$-8 sources in the HFF observations are star
cluster complexes commonly located within star-forming galaxies at
$z=0$ (Kennicutt et al.\ 2003; Bastian et al.\ 2005) and which can be
seen out to $z\sim3$ in strongly lensed galaxies (Jones et al.\ 2010;
Wisnioski et al.\ 2012; Livermore et al.\ 2012, 2015; Swinbank et
al.\ 2012; Adamo et al.\ 2013; Vanzella et al.\ 2017b;
Dessauges-Zavadsky et al.\ 2017).  

Star cluster complexes -- often referred to as cluster complexes in
nearby galaxies -- are known to show a range of surface brightnesses
at all redshifts where they are observed, i.e., $z\sim0$-3 (Bastian et
al.\ 2005, 2006; Jones et al.\ 2010; Swinbank et al. 2012; Wisnioski
et al.\ 2012; Rodr{\'{\i}}guez-Zaur{\'{\i}}n et al.\ 2011; Kennicutt
et al.\ 2003).  Star cluster complexes are also described as
star-forming clumps (or giant HII regions) when observed in distant
galaxies.  A simple fit to the mean surface brightness of star cluster
complexes as function of redshift yields the following relation
(Livermore et al.\ 2015):
\begin{equation}
\log \left( \frac{\Sigma_{clump}}{M_{\odot}\, \textrm{yr}^{-1}\, \textrm{kpc}^{-2}} \right) = (3.5\pm0.5)\log(1+z)-(1.7\pm0.2)
\label{eq:sscevol}
\end{equation}
While many other observations of star cluster complexes at
intermediate to high redshifts are also consistent with the above
trend (Franx et al.\ 1997; Swinbank et al.\ 2009; Wuyts et al.\ 2014;
Johnson et al.\ 2017), some star cluster complexes at $z\sim0$ have
been reported to show much higher (by factors of $\sim$100) surface
densities of star formation (Fisher et al.\ 2017).

The implied evolution in the surface brightness of star cluster
complexes is essentially identical to what one would infer from
dimensional arguments.  The sizes of collapsed sources is generally
found to scale as $(1+z)^{-1}$ (e.g., Bouwens et al.\ 2004; Oesch et
al.\ 2010; Ono et al.\ 2013; Holwerda et al.\ 2015; Shibuya et
al.\ 2015) and the evolution in dynamical time goes as $(1+z)^{-1.5}$,
such that $\Sigma_{SFR} \propto r^{-2} t_{dyn} ^{-1}$ $\propto
(1+z)^{3.5}$.  Nevertheless, it should be recognized that the best-fit
evolution in $\Sigma_{clump}$ with redshift likely suffers from
surface brightness selection effects (as only the highest surface
brightness star cluster complexes can be identified at a given
redshift), so the evolution suggested by Eq.~\ref{eq:sscevol} should
only be considered indicative.

We include a gray-shaded trapezoid in Figure~\ref{fig:msre00} to show
the region in size-luminosity parameter space star cluster complexes
in $z\sim0$-3 galaxies have been found to inhabit.  The light gray
region shows an extrapolation of this relation to $z=3$-6.  The solid
red circles correspond to the measured sizes and equivalent $UV$
luminosities of the star cluster complexes identified in the highly
magnified $z=4.92$ galaxy behind MS1358+62 (Franx et al.\ 1997) by
Swinbank et al.\ (2009) and Jones et al.\ (2010), while the solid cyan
circle and green triangle correspond to the sizes and luminosities of
IIZw40 and 30 Doradus, respectively (English \& Freeman 2003; Vanzi et
al.\ 2008).

Even though we present star cluster complexes at a given redshift as
having a fixed surface brightness, Wisnioski et al.\ (2012) have found
their size $r$ to vary as $L^{1/(2.72\pm0.04)} \sim L^{0.37\pm0.01}$
where $L$ correspond to the H$\alpha$ luminosities, such that the most
luminous star cluster complexes also had the highest surface brightnesses.

\subsection{Size-Luminosity Relation for Star Clusters and Super Star Clusters}

Finally, as a third reference point, we consider the region in
parameter space occupied by star clusters and super star clusters.
Not attempt here will be done to summarize the substantial work has
been done on this topic (e.g., Meurer et al.\ 1995; Rejkuba et
al.\ 2007; Murray 2009; Bastian et al.\ 2013), but only to indicate
where star clusters lie in parameter space.

Meurer et al.\ (1995) provide a convenient summary of where star
clusters lie in terms of their effective radii and $UV$ luminosities
$M_{UV}$ in their Figure 14.  The purple line in the right panel of
Figure~\ref{fig:msre00} demarcates the approximate region in parameter
space that star clusters and super star clusters populate.  $UV$
luminosities of the star clusters extend from $-$9 mag to $-19$ mag,
masses range from $10^4$ to $10^8$ $M_{\odot}$ (Maraston et al.\ 2004;
Cabrera-Ziri et al.\ 2014, 2016), while the typical effective radii of
star clusters range from 0.5 pc to 4 pc (e.g., Lada \& Lada 2003).

The most massive ($>$10$^5$ $M_{\odot}$) star clusters are often
called super star clusters, with the effective radii extend up to
$\sim$20 pc (e.g., Meurer et al.\ 1995; Rejkuba et al.\ 2007; Murray
2009; Bastian et al.\ 2013).  Meurer et al.\ (1995) classify any star
clusters with $UV$ luminosities greater than $-$14 mag as super star
clusters.

\section{Sizes of $z\geq6$ HFF sources} 

\subsection{Measurement Procedure}

In fitting the two-dimensional spatial profile of galaxies behind the
HFF clusters to measure sizes, we must account for the substantial
impact that gravitational lensing from the foreground cluster has on
the spatial profile of galaxies.  

The relevant quantities in computing the size of a lensed source is
both the total magnification factor $\mu$ and the source shear.  In
Bouwens et al.\ (2017a), we introduced a quantity that we called the
shear factor $S$ which we defined as follows:
\begin{equation}
S = \left\{
\begin{array}{lr}
\frac{1-\kappa - \gamma}{1-\kappa+\gamma}, &
\text{for } \frac{1-\kappa - \gamma}{1-\kappa+\gamma} \geq 1\\
\frac{1-\kappa + \gamma}{1-\kappa-\gamma}, &
\text{for } \frac{1-\kappa - \gamma}{1-\kappa+\gamma} < 1
\end{array}\right.
\end{equation}
where $\kappa$ is the convergence and $\gamma$ is the shear.  The
shear factor $S$ gives the axis ratio a circular galaxy would have due
to the impact of gravitational lensing.

The source magnification $\mu$ can be computed from the convergence
$\kappa$ and shear $\gamma$ maps:
\begin{displaymath}
\mu = \frac{1}{(1-\kappa)^2 - \gamma^2}
\end{displaymath}

The impact of the gravitational lensing on background galaxies is to
stretch sources by the factor $\mu^{1/2} S^{1/2}$ along the major
shear axis and by the factor $\mu^{1/2} S^{-1/2}$ perpendicular to the
major shear axis.  

\begin{figure}
\epsscale{1.17}
\plotone{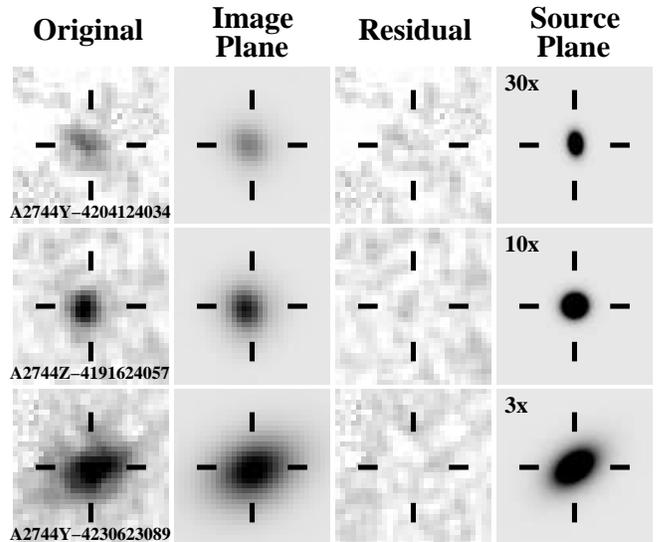}
\caption{Illustration of the typical profile fits used here (\S3.1) in
  deriving half-light radii for sources lensed by the HFF clusters.
  The source-plane model profiles in the rightmost column
  (\textit{shown at the various zoom factors indicated in the
    rightmost postage stamp}) are transformed into the image plane and
  convolved with the PSF to produce the model profiles in the image
  plane (\textit{shown in the second leftmost column}) for comparison
  with the observed two-dimensional profile (\textit{leftmost
    column}), which is an inverse-weighted mean coaddition of the
  $Y_{105}$, $J_{125}$, $JH_{140}$, and $H_{160}$ images.  The
  residuals of our profile fits are shown in the second rightmost
  column.\label{fig:fit}}
\end{figure}

We estimate the half-light radii of sources via a Markov chain
Monte-Carlo (MCMC) algorithm where we compare the observed
two-dimensional profile with a lensed model profile of a model source
with a Sersic radial profile with major and minor axes oriented at
some position angle on the sky.  In fitting to the two dimensional
profile, we coadd the $Y_{105}$, $J_{125}$, $JH_{140}$, and $H_{160}$
images together after scaling to the fluxes in the images to a fixed
$f_{\nu}$ frequency and weighting the images by the inverse variance.
We coadd the $Y_{105}$, $J_{125}$, $JH_{140}$, and $H_{160}$ PSFs in
the same way to derive a composite PSF for the fit procedure.  We fix
the Sersic parameter to 1, but find similar results (albeit slightly
larger sizes by a factor of 1.5) using other Sersic parameters ($n=2$,
3).  Lensing is modeled as magnifying the source by the factor
$\mu^{1/2} S^{1/2}$ along the major shear axis and by the factor
$\mu^{1/2} S^{-1/2}$ along the minor shear axis.

\begin{deluxetable}{ccc}
\tablecolumns{3}
\tabletypesize{\footnotesize}
\tablecaption{Parametric Lensing Models Utilized (see also \S4.1)\tablenotemark{a}\label{tab:models}}
\tablehead{\colhead{Cluster} & \colhead{Model} & \colhead{Version}}
\startdata
Abell 2744 & CATS & v4.1 \\
           & Sharon/Johnson & v4 \\
           & Keeton & v4 \\
           & GLAFIC & v3 \\
           & Zitrin/NFW & v3 \\\\

MACS0416 & CATS & v4.1 \\
         & Sharon/Johnson & v4 \\
         & Keeton & v4 \\
         & GLAFIC & v3 \\
         & Zitrin/NFW & v3 \\
         & Caminha & v4 \\\\

MACS0717 & CATS & v4.1 \\
         & Sharon/Johnson & v4 \\
         & Keeton & v4 \\
         & GLAFIC & v3 \\\\

MACS1149 & CATS & v4.1 \\
         & Sharon/Johnson & v4 \\
         & Keeton & v4 \\
         & GLAFIC & v3
\enddata
\tablenotetext{a}{We only make use of medians of post-HFF parametric lensing 
models to represent the lensing magnification of sources behind the HFF clusters.
Even so, we emphasize that the non-parametric lensing models (\textsc{Grale}:
Liesenborgs et al.\ 2006; Sebesta et al.\ 2016, Brada{\v c}: Brada{\v c} et
al.\ 2009; Hoag et al.\ 2017, Zitrin-LTM: Zitrin et al.\ 2012, 2015,
Diego: Lam et al.\ 2014; Diego et al.\ 2015a, 2015b, 2016a, 2016b,
2017) also perform very well.}
\end{deluxetable}

Figure~\ref{fig:fit} illustrates our two-dimensional profile fits for
three sources in our catalogs, showing the original images
(\textit{leftmost column}), the PSF-convolved model images in the
image plane (\textit{second leftmost column}), the residual image
(\textit{second rightmost column}), and finally the zoomed model
images in the source plane before PSF convolution (\textit{rightmost
  column}).

We now describe the magnification factors $\mu$ and shear factors $S$
that we utilize in our analysis.  For the sake of robustness, we do
not rely on the results from a single lensing model -- since lensing
models lack predictive power when the magnification factors from the
models become particularly high, as we illustrate for the linear
magnification factor $\mu^{1/2} S^{1/2}$ in Appendix A and previously
demonstrated in Bouwens et al.\ (2017b) for the magnification factor.
One can do better using the median model (Bouwens et al.\ 2017b;
Livermore et al.\ 2017).  

\begin{figure}
\epsscale{1.17}
\plotone{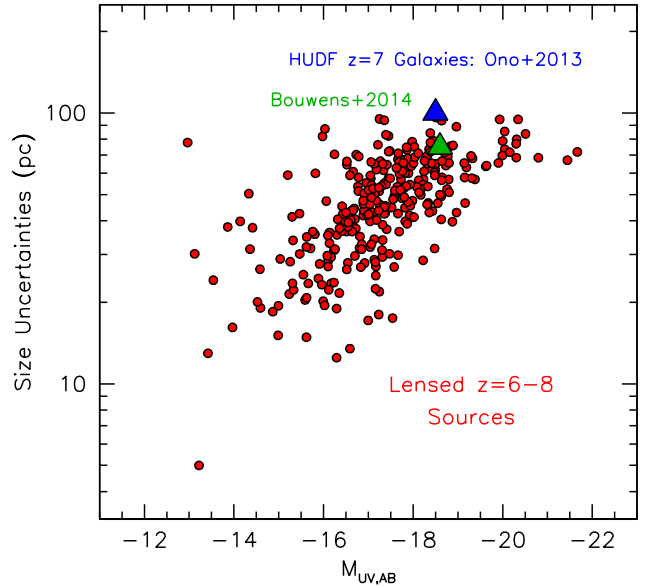}
\caption{Nominal $1\sigma$ accuracy with which source sizes can be
  measured for individually lensed $z=6$-8 sources identified behind
  various HFF clusters vs. the $UV$ luminosity inferred (\textit{black
    crosses}).  The accuracy of size measurements is computed by
  adding in quadrature the size uncertainty based on the MCMC fit
  results and the size uncertainty resulting from the unknown lensing
  magnification (based on the dispersion in the lensing models). For
  comparison, we also show the accuracy with which size measurements
  were claimed for individual $z=7$-8 sources from the HUDF data (Ono
  et al.\ 2013).  Bouwens et al.\ (2014) estimated a $1\sigma$
  accuracy of 75 pc based on their analysis of sizes in a stacked
  $z\sim7$ source from the HUDF.\label{fig:resol}}
\end{figure}

We therefore take the median magnification and shear factors from all
available parametric lens models, including CATS (Jullo \& Kneib 2009;
Richard et al.\ 2014; Jauzac et al.\ 2015a,b; Limousin et al.\ 2016;
Mahler et al.\ 2017; Lagattuta et al.\ 2017), Sharon/Johnson (Johnson
et al.\ 2014), GLAFIC (Oguri 2010; Ishigaki et al.\ 2015; Kawamata et
al.\ 2016), Zitrin-NFW (Zitrin et al.\ 2013, 2015), Keeton (Keeton
2010), and Caminha et al.\ (2016, 2017).  Each of the four clusters we
utilize have highly-refined models available for most but typically
not all varieties of model.  Our Abell 2744 median model makes use of
5 of the models (v4.1 of CATS, v4 of Sharon/Johnson, v3 of GLAFIC, v3
of Zitrin-NFW, v4 of Keeton), our MACS0416 median model makes use of 6
of the models (v4.1 of CATS, v4 of Sharon/Johnson, v3 of GLAFIC, v3 of
Zitrin-NFW, v4 of Keeton, v4 of Caminha), while our MACS0717 and
MACS1149 median models make use of 4 of the models (v4.1 of CATS, v4
of Sharon/Johnson, v3 of GLAFIC, v4 of Keeton).
Table~\ref{tab:models} provides a convenient summary of the models we
use.

\begin{deluxetable*}{ccccccc}
\tablecolumns{7}
\tabletypesize{\footnotesize}
\tablecaption{Catalog of Tiny Star-Forming Sources\tablenotemark{a} at $z\sim6$-8\label{tab:tiny}}
\tablehead{\colhead{ID} & \colhead{R.A.} & \colhead{Decl} & \colhead{$M_{UV}$} & \colhead{$\mu$\tablenotemark{b}} & \colhead{$\mu_{1D}$\tablenotemark{c}} & \colhead{$r_e$ (pc)\tablenotemark{d}}}
\startdata
A2744I-4205324088*\tablenotemark{e} &  00:14:20.54  &  $-$30:24:08.9  & $-$14.6$_{-0.6}^{+0.6}$  & 21.5$_{-8.8}^{+15.3}$  & 6.2$_{-2.5}^{+1.9}$  & 17$_{-13}^{+28}$\\
A2744I-4222023578  &  00:14:22.21  &  $-$30:23:57.9  & $-$15.6$_{-1.2}^{+0.7}$  & 50.1$_{-33.6}^{+41.4}$  & 8.5$_{-3.8}^{+4.4}$  & 31$_{-13}^{+32}$\\
A2744I-4212723104  &  00:14:21.28  &  $-$30:23:10.5  & $-$16.0$_{-0.1}^{+0.1}$  & 6.4$_{-0.8}^{+0.5}$  & 5.7$_{-1.2}^{+0.4}$  & 32$_{-17}^{+24}$\\
A2744Y-4204124034*  &  00:14:20.41  &  $-$30:24:03.5  & $-$14.0$_{-1.6}^{+1.2}$  & 75.0$_{-58.0}^{+151.2}$  & 15.1$_{-9.5}^{+24.0}$  & 14$_{-9}^{+29}$\\
M0416I-6055105026  &  04:16:05.52  &  $-$24:05:02.7  & $-$14.6$_{-0.4}^{+1.4}$  & 18.0$_{-6.1}^{+47.0}$  & 16.3$_{-6.1}^{+35.8}$  & 40$_{-27}^{+26}$\\
M0416I-6090604399  &  04:16:09.06  &  $-$24:04:40.0  & $-$15.3$_{-0.5}^{+0.3}$  & 11.6$_{-4.6}^{+3.4}$  & 7.3$_{-3.4}^{+2.4}$  & 32$_{-13}^{+38}$\\
M0416I-6095704260*  &  04:16:09.57  &  $-$24:04:26.1  & $-$15.0$_{-0.2}^{+0.3}$  & 12.5$_{-2.5}^{+3.3}$  & 8.0$_{-2.0}^{+2.5}$  & 28$_{-12}^{+19}$\\
M0416I-6120203507  &  04:16:12.02  &  $-$24:03:50.8  & $-$13.6$_{-1.4}^{+1.2}$  & 62.7$_{-45.6}^{+121.6}$  & 57.6$_{-42.7}^{+91.5}$  & 18$_{-11}^{+53}$\\
M0416I-6118103480*\tablenotemark{$\ddagger$}  &  04:16:11.81  &  $-$24:03:48.1  & $-$15.0$_{-1.0}^{+1.3}$  & 33.6$_{-19.9}^{+81.6}$  & 24.5$_{-15.2}^{+85.1}$  & 16$_{-13}^{+29}$\\
M0416I-6130803432\tablenotemark{$\ddagger$}  &  04:16:13.09  &  $-$24:03:43.3  & $-$17.3$_{-0.2}^{+0.2}$  & 3.5$_{-0.6}^{+0.7}$  & 3.4$_{-0.6}^{+0.3}$  & 38$_{-27}^{+36}$\\
M0416I-6115434445\tablenotemark{$\dagger$,$\ddagger$}  &  04:16:11.54  &  $-$24:03:44.5  & $-$14.5$_{-1.0}^{+0.9}$  & 43.0$_{-25.4}^{+54.2}$  & 31.0$_{-18.4}^{+48.0}$  & 21$_{-13}^{+31}$\\
M0416I-6106703335*  &  04:16:10.67  &  $-$24:03:33.6  & $-$16.3$_{-0.2}^{+0.3}$  & 11.1$_{-2.0}^{+3.7}$  & 8.8$_{-0.7}^{+3.1}$  & 33$_{-12}^{+13}$\\
M0416I-6114803434\tablenotemark{$\dagger$,$\ddagger$}  &  04:16:11.48  &  $-$24:03:43.5  & $-$17.0$_{-0.5}^{+0.3}$  & 21.3$_{-7.6}^{+6.3}$  & 14.9$_{-4.8}^{+7.8}$  & 38$_{-14}^{+21}$\\
M0416Y-6088104378*  &  04:16:08.82  &  $-$24:04:37.9  & $-$13.4$_{-1.1}^{+1.0}$  & 62.1$_{-39.5}^{+90.8}$  & 38.3$_{-27.0}^{+45.7}$  & 11$_{-6}^{+28}$\\
M0717I-7354743496  &  07:17:35.48  &  37:43:49.7  & $-$13.9$_{-1.4}^{+1.6}$  & 57.7$_{-42.1}^{+205.4}$  & 46.8$_{-34.0}^{+142.4}$  & 27$_{-20}^{+72}$\\
M0717I-7374244282  &  07:17:37.43  &  37:44:28.3  & $-$15.0$_{-0.8}^{+0.8}$  & 9.6$_{-5.0}^{+10.8}$  & 3.6$_{-0.9}^{+1.5}$  & 34$_{-22}^{+38}$\\
M0717I-7361844009*  &  07:17:36.18  &  37:44:01.0  & $-$13.2$_{-0.4}^{+0.3}$  & 67.2$_{-22.1}^{+18.6}$  & 22.2$_{-1.2}^{+3.6}$  & 23$_{-5}^{+5}$\\
M0717I-7357345028  &  07:17:35.74  &  37:45:02.9  & $-$14.4$_{-1.6}^{+1.2}$  & 26.8$_{-20.5}^{+54.2}$  & 19.0$_{-14.5}^{+58.5}$  & 24$_{-18}^{+79}$\\
M0717Z-7390844017\tablenotemark{$\ddagger$}  &  07:17:39.09  &  37:44:01.8  & $-$17.1$_{-1.8}^{+0.4}$  & 21.8$_{-17.8}^{+10.7}$  & 17.7$_{-13.3}^{+14.4}$  & 24$_{-11}^{+75}$\\
M0717Z-7401344384  &  07:17:40.14  &  37:44:38.5  & $-$16.1$_{-0.7}^{+0.0}$  & 8.3$_{-3.9}^{+0.3}$  & 6.4$_{-2.3}^{+1.5}$  & 39$_{-15}^{+33}$\\
M0717Z-7311744437  &  07:17:31.18  &  37:44:43.8  & $-$16.0$_{-0.4}^{+0.8}$  & 8.2$_{-2.6}^{+8.8}$  & 4.0$_{-0.7}^{+3.1}$  & 27$_{-18}^{+30}$\\
M0717Y-7336744331  &  07:17:33.68  &  37:44:33.2  & $-$14.4$_{-1.3}^{+2.3}$  & 43.1$_{-29.9}^{+312.6}$  & 7.6$_{-3.7}^{+40.9}$  & 31$_{-26}^{+38}$\\
M0717Y-7329744137  &  07:17:32.97  &  37:44:13.8  & $-$15.3$_{-0.6}^{+0.9}$  & 13.7$_{-5.9}^{+16.8}$  & 7.5$_{-3.7}^{+8.6}$  & 34$_{-19}^{+42}$\\
M1149I-9384023344  &  11:49:38.40  &  22:23:34.5  & $-$13.0$_{-2.4}^{+2.6}$  & 124.7$_{-111.1}^{+1226.2}$  & 90.7$_{-81.8}^{+949.9}$  & 26$_{-24}^{+252}$\\
M1149I-9379223320  &  11:49:37.93  &  22:23:32.1  & $-$13.1$_{-1.2}^{+0.4}$  & 74.7$_{-49.4}^{+37.3}$  & 39.8$_{-24.6}^{+31.6}$  & 35$_{-16}^{+57}$\\
M1149Y-9377423253*  &  11:49:37.74  &  22:23:25.4  & $-$14.9$_{-0.3}^{+0.1}$  & 16.2$_{-4.1}^{+2.0}$  & 7.9$_{-3.9}^{+1.3}$  & 19$_{-11}^{+31}$
\enddata
\tablenotetext{a}{All sources with inferred half-light radii less than 40 pc are included in this table.  Star forming sources could include star cluster complexes, super star clusters, proto-globular clusters, or especially compact galaxies}
\tablenotetext{b}{Median magnification factors (and $1\sigma$ uncertainties) derived weighting equally the latest public version 3/4 parametric models from each lensing methodology (\S4.1).}
\tablenotetext{c}{$\mu_{1D}$ are the median one-dimensional magnification factors (and $1\sigma$ uncertainties) along the major shear axis $\mu^{1/2} S^{1/2}$ weighting equally the parametric models from each lensing methodology.  This is the same quantity as $\mu_{tang}$ reported by Vanzella et al.\ (2017a).}
\tablenotetext{d}{Inferred half-light radius in physical units.  The quoted uncertainties include both uncertainties in the spatial fits and uncertainties in the lensing model.}
\tablenotetext{e}{The eight sources where the upper $1\sigma$ limit on the inferred half-light radius is less than 50 pc are marked with an ``*.''}
\tablenotetext{$\dagger$}{Tiny star-forming source also presented in Vanzella et al.\ (2017a).}
\tablenotetext{$\ddagger$}{Source also has an inferred size of $\leq$40 pc in  the Kawamata et al.\ (2017) catalog.}
\end{deluxetable*}

\begin{figure*}
\epsscale{1.08}
\plotone{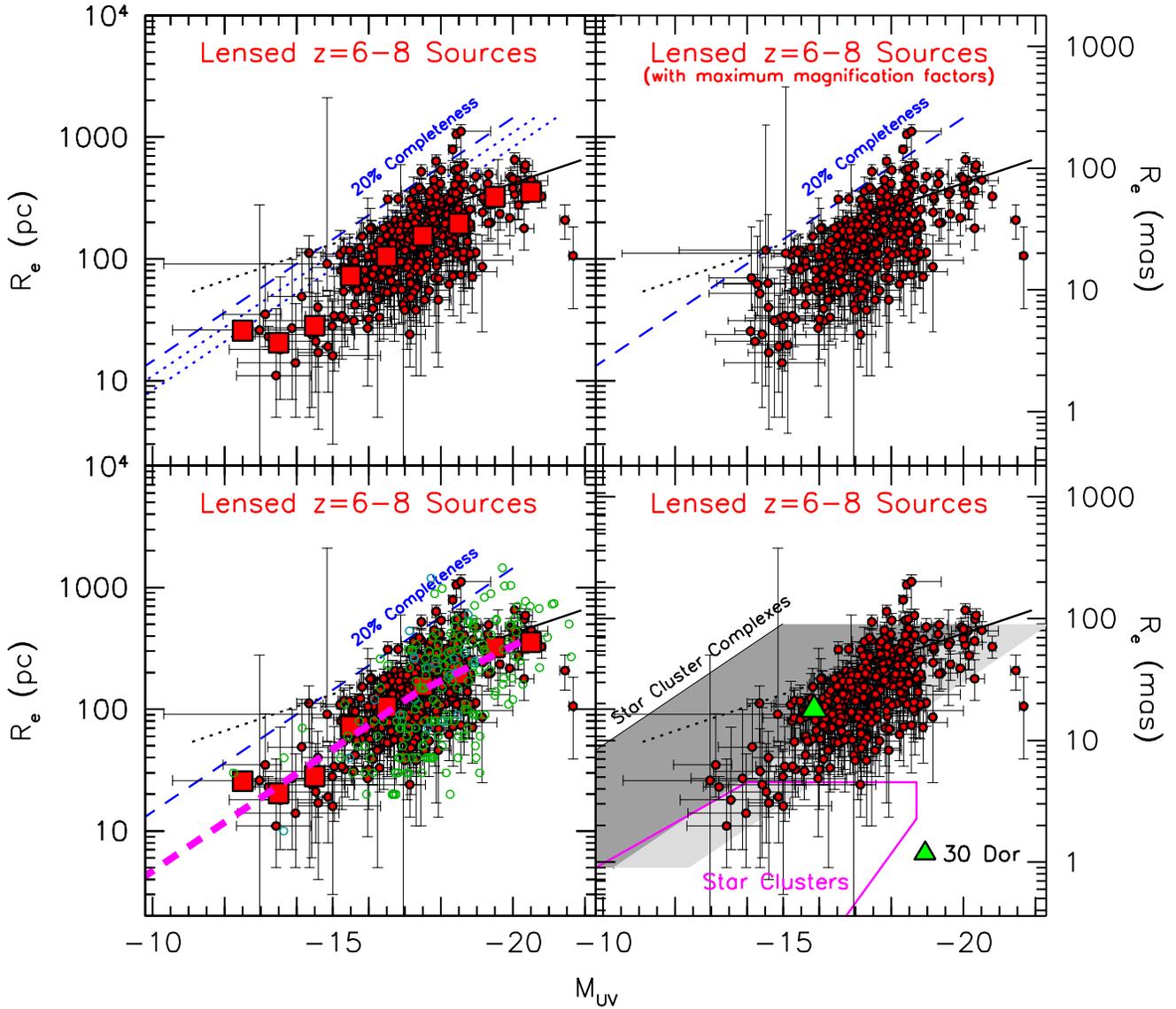}
\caption{Comparison of the distribution of sizes and luminosities for
  $z=6$-8 galaxies in the HFFs with a canonical size-luminosity
  relationship from blank field studies (\textit{black solid and
    dotted line in all panels}) and those of star cluster complexes
  (\textit{lower right panel}).  The canonical size-luminosity
  relation for galaxies and star cluster complexes is as in
  Figure~\ref{fig:msre00}.  For the sizes and luminosities of
  individual lensed $z=6$-8 galaxies, the results (\textit{red
    circles} and $1\sigma$ limits) are based on the median
  magnification factors from the parametric models (\textit{upper
    left, lower left, and lower right panels}) and where we impose a
  maximum linear magnification factor of 20 (\textit{upper right
    panel}) for greater robustness.  $1\sigma$ errors on the inferred
  sizes and $UV$ luminosities are quoted based on the 68\% confidence
  intervals in the size fits and the range of magnification factors in
  the parametric lensing models.  The large red squares indicate the
  median measured half-light radius per 1-mag $UV$ luminosity bin.
  The blue dashed line delimits the region where our selections are
  expected to be less than 20\% complete.  The blue dotted lines
  directly to the right of the blue dashed lines delimit the regions
  where our selections are expected to be 50\% and 80\% complete
  relative to the maximum.  As these completeness limits crosses the
  standard size-luminosity relation at $\sim$$-$15 mag, we might
  expect selections in the HFFs to be significantly incomplete at
  $>$$-$15 mag, if this relation applied to the lowest luminosity
  $z=6$-8 galaxies.  If our selections are largely complete, the
  dashed magenta line (\textit{lower left panel}) shows the asymptotic
  form for the size-luminosity relation (where radius $\propto$
  $L^{0.5}$).  Similar trends are seen in the size vs. luminosity for
  lower luminosity ($M_{UV,AB}>-19$) galaxies by Kawamata et
  al.\ (2017: \textit{light green open circles}) and Laporte et
  al.\ (2016: \textit{light open cyan circles}).  Faint $z\sim6$-8
  galaxies also exhibit very similar sizes to that seen for star
  cluster complexes like 30 Doradus at $z\sim0$ (\textit{solid green
    triangle}) and some star cluster complexes at $z\geq 2$
  (\textit{lower right panel}).  The magenta lines demarcate the star
  cluster region as shown in
  Figure~\ref{fig:msre00}.\label{fig:msre0}}
\end{figure*}

In general, the parametric lens models appear to perform slightly
better in terms of their predictive power than the non-parametric
models (Meneghetti et al.\ 2017) though the non-parametric models
(\textsc{Grale}: Liesenborgs et al.\ 2006; Sebesta et al.\ 2016,
Brada{\v c}: Brada{\v c} et al.\ 2009; Hoag et al.\ 2017, Zitrin-LTM: Zitrin et
al.\ 2012, 2015, Diego: Lam et al.\ 2014; Diego et al.\ 2015a, 2015b,
2016a, 2016b, 2017) also do very well.

In computing the magnification and shear factors for the individual
models (to produce the median), we multiply the relevant $\kappa$ and
$\gamma$'s from the aforementioned public models by the ratio of the
distance moduli $D_{ls}/D_{s}$, where $D_{ls}$ is the angular diameter
distance between the lensing cluster and source and the angular
diameter distance to the source, using the best-fit photometric
redshift for the source to compute the distance.  

In this way, we compute the median linear magnification factor
$\mu^{1/2} S^{1/2}$ and $\mu^{1/2} S^{-1/2}$ along the major and minor
shear axes, respectively.  It is worth remarking that these linear
magnification factors appear to be reliable to values as high as 20,
if we take the results of Appendix A as indicative, but not in excess
of 20.  The direction of the major shear axis is derived using the
version 4.1 CATS magnification model, but is fairly similar for the
other parametric lensing models.

It is interesting to ask how well we can use the HFF lensing clusters
to determine the scale length of faint galaxies to very small sizes.
We can look to some recent work from HST imaging observations over the
Hubble Ultra Deep Field (Beckwith et al.\ 2006; Bouwens et al.\ 2011;
Ellis et al.\ 2013; Illingworth et al.\ 2013) to provide some
indication.  Ono et al.\ (2013) measure source sizes for $z\sim7$-8
galaxies at $\sim$$-$19 mag to a $1\sigma$ uncertainty of $\sim$100 pc
and at $\sim$$-$18 mag to a $1\sigma$ uncertainty of $\sim$150 pc,
corresponding to $\sim$0.1 native pixel length.  In Bouwens et
al.\ (2014), the sizes of a stack of $z\sim7$ galaxies are measured to
an estimated $1\sigma$ accuracy of 75 pc at $\sim-$18.5.  If we assume
that the median linear magnification factors are accurate to factors
of 20, this means we can measure source sizes to 20$\times$ higher
spatial resolution over the HFF clusters as we can over the Hubble
Ultra Deep Field.  This means we can potentially measure the linear
sizes of sources to a $1\sigma$ accuracy of 4-5 pc.

In Figure~\ref{fig:resol}, we provide a sense for the accuracies with
which we can measure sizes for our lensed $z=6$-8 samples vs. $UV$
luminosity.  The accuracy of size measurements is computed by adding
in quadrature the size uncertainty based on the MCMC fit results and
the size uncertainty resulting from the unknown lensing magnification
(based on the dispersion in the lensing models).  This suggests a
typical half-light radius measurement accuracy of 50 pc and 10 pc for
sources at $-18$ mag and $-$15 mag, respectively.

\subsection{Size vs. Luminosity Results}

In the upper left panel of Figure~\ref{fig:msre0}, we show the
measured sizes and estimated luminosities of lensed sources in our
$z=6$-8 samples in relation to the derived and extrapolated
size-luminosity relation from blank field studies.  We also indicate
in this panel where our source recovery experiments from Bouwens et
al.\ (2017a, 2017b) indicate that our selection would be less than
20\% complete (\textit{blue dashed line}).

Given uncertainties in our knowledge of the magnification factors for
specific lensed sources behind the HFF clusters, we also present in
the upper right panel of Figure~\ref{fig:msre0} the source sizes and
luminosities, capping the linear magnification factor and
magnification factors to values of 20 and 30, respectively.  The
position of sources in parameter space is similar to the upper left
panel, but with fewer sources at especially small sizes and low
luminosities.

Interestingly enough, at the bright end ($<$$-$17 mag), lensed sources
in our samples scatter around extrapolated expected sizes from
blank-field studies, with a best-fit trend such that half-light radius
scales with luminosity as $L^{0.26\pm0.03}$.  However, as one
considers the size-luminosity relation defined by our sample to lower
luminosities, one sees a trend to very small sizes with an asymptotic
slope radius $\propto$ $L^{0.50_{-0.11}^{+0.10}}$ where $L$ is luminosity
(\textit{magenta dashed line}):
\begin{displaymath}
r_e = (72_{-5}^{+6}\, \textrm{pc})10^{-0.4(M_{UV}+16)(0.50_{-0.11}^{+0.10})}
\end{displaymath}
The measured size-luminosity relation derived by Kawamata et
al.\ (2017: \textit{light green open circles}) and Laporte et
al.\ (2016: \textit{cyan open circles}) for lower-luminosity galaxies
follows a similar trend.  In deriving the radius vs. luminosity
relation, the intrinsic scatter found by Shibuya et al.\ (2015), i.e.,
0.2 dex, is added in quadrature to the measurement errors.  In Bouwens
et al.\ (2017a) and Kawamata et al.\ (2017), the reported dependence
of size on luminosity are $\propto L^{0.50\pm0.07}$ and $\propto
L^{0.46_{-0.09}^{+0.08}}$, respectively.

\begin{figure*}
\epsscale{1.14}
\plotone{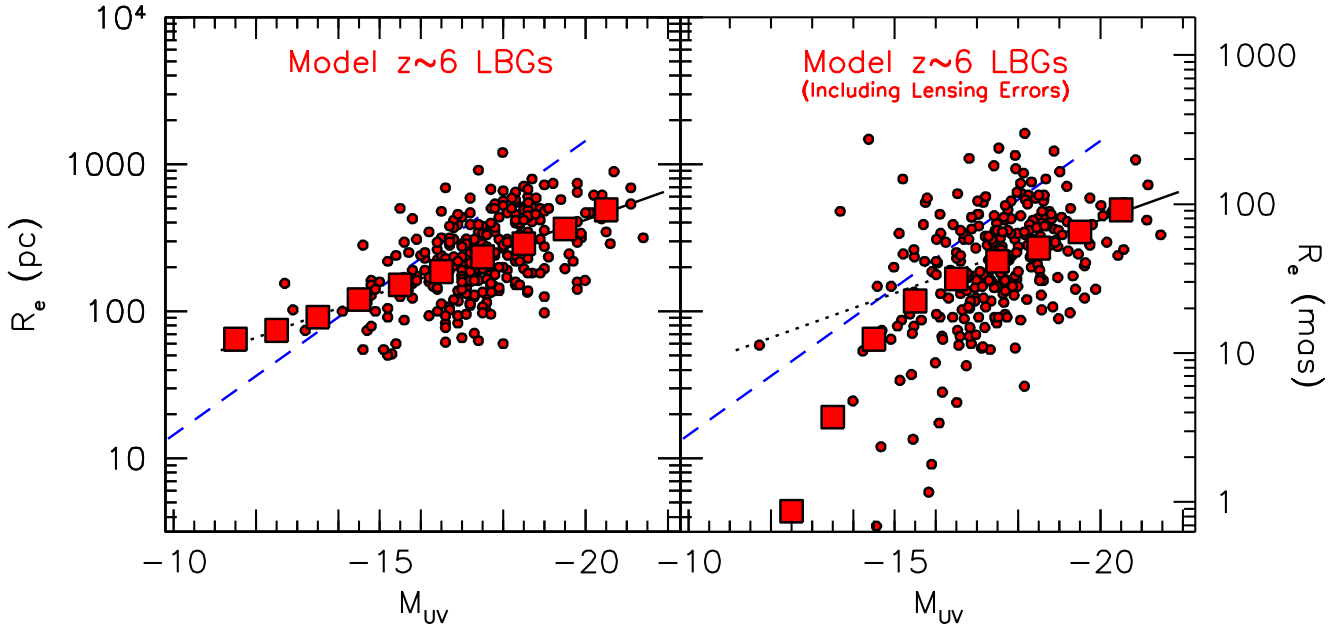}
\caption{Illustration of how errors in the lensing models are expected
  to impact the size-luminosity distribution using Monte-Carlo
  simulations.  The left panel shows an input distribution of sizes
  and luminosities for a $\sim$300 source sample (\textit{shown with
    the red solid points}) using the size-luminosity relation from the
  Shibuya et al.\ (2015).  The right panel shows the recovered
  distribution of sizes and luminosities derived treating the
  \textsc{CATS} models as the truth and using the median magnification
  and shear maps from the parametric lensing models.  The red squares
  show the median size at a given $UV$ luminosity -- with the median
  shown with the red squares -- before and after applying the median
  magnification and shear maps.  The red median sizes are derived
  based on 100$\times$ more sources than are shown on the figure.  No
  selection effects are included in these idealized results.  Errors
  in the lensing models can introduce significant scatter in the
  recovered sizes and $UV$ luminosities, but do not appreciably impact
  the recovered median size measurements brightward of $-$15 mag (but
  do so at fainter luminosities, see \S4.3).  Nevertheless, we find a
  slightly larger number of small $<$40-pc sources in the
  observations, i.e., 26, than in these simulations (typically 15),
  suggesting that some of the small sources are likely
  bona-fide.\label{fig:lenserror}}
\end{figure*}

There are reasons for supposing that the observed trend might arise
due to surface brightness selection effects, uncertainties in the
lensing models, or a combination of the two effects.  Indeed, surface
brightness selection effects (e.g., Bouwens et al.\ 2004; Oesch et
al.\ 2015; Taghizadeh-Popp et al.\ 2015) might cause us to recover a
$r\propto L^{0.50_{-0.11}^{+0.10}}$ correlation between size and luminosity, as
was pointed out in both Bouwens et al.\ 2017a and Ma et al.\ 2017.
The dashed line in Figure~\ref{fig:msre0} show the sizes and
luminosities where source selection fraction is only 20\% efficient
(vs. the maximum) using the simulations in Bouwens et al.\ (2017a,
2017b), while the dotted lines in the upper left panel of
Figure~\ref{fig:msre0} show the sizes and luminosities where the
selection fraction is 50\% and 80\%.  Such a correlation between
source size and luminosity would only be enhanced by uncertainties in
the lensing model (scattering sources along the same general
radius-luminosity vector).

It is obviously useful to examine the form of the size luminosity
relation to very luminosities in a way that are robust against such
worries.  In principle, such is possible relying on the faintest
$z\sim4$ sources identified over the HUDF.  In appendix B, we provide
an independent measurement of the median sizes of $z\sim4$ galaxies as
a function of the $UV$ luminosity, and compare the median sizes with
what we derive from a faint $z\sim4$ selection behind the HFF
clusters.

Beyond the plausibility tests we provide on our size-luminosity
measurements in Appendix C, there are two other arguments we can make
which provide some support to our overall constraints on the median
measured sizes of galaxies vs. $UV$ luminosity.  This first argument
(\S4.3) relies on the impact the size distribution has on the form of
the $UV$ LF at $>$$-$15 mag (\S4.3) and the second argument (\S4.4)
relies on simulations designed to estimate the impact of the model
uncertainties on the number of small sources recovered in our HFF
samples.

Assuming that these arguments are valid (and the results in Appendix C
are indicative), our results indicate a break in the size-luminosity
relation at $M_{UV}\sim-17$ mag, such that the sizes of galaxies
transition from lying along the size-luminosity relation of more
luminous galaxies to possessing sizes and luminosities more similar to
star cluster complexes in $z=0$-3 galaxies (Bastian et al.\ 2006;
Jones et al.\ 2010; Livermore et al.\ 2012, 2015; Wisnioski et
al.\ 2012; Swinbank et al.\ 2012; Johnson et al.\ 2017), as explicitly
shown in the lower right panel of Figure~\ref{fig:msre0}.  In fact,
the typical $-$15 mag galaxy in our samples has a smaller half-light
radius than 30 Doradus, which has a measured half-light radius of
$\sim$100 pc (\textit{lower right panel of Figure~\ref{fig:msre0}}).

It has been suggested that some lensed high-redshift sources behind
the HFF clusters may in fact be super star clusters (Vanzella et
al.\ 2017a, 2017b; Bouwens et al.\ 2017a [\S6.1]).  It is interesting
therefore to ask if any sources from our samples seem consistent with
corresponding to super star clusters.  In Table~\ref{tab:tiny} we
provide such a compilation, including all sources with estimated
half-light radii that could plausibly correspond to super star
clusters.  As sizes of super star clusters from 4 pc to 20 pc, we
include sources with estimated sizes up to 40 pc.

The uncertainties we report on the measured sizes include a $1\sigma$
error computed based on the range in linear magnifications predicted
by the parametric lensing models.  The uncertainties in the measured
sizes for sources in Table~\ref{tab:tiny} are substantial.  Only two
of the sources have measured sizes less than 40 pc after allowing for
the $1\sigma$ uncertainties in the measured sizes and from the lensing
models.  Eight of the sources are within an upper bound of 50 pc (a
slightly less stringent limit) allowing for the $1\sigma$
uncertainties.  These sources are indicated in Table~\ref{tab:tiny}
with a ``*''.

\subsection{Impact of Uncertainties in the Lensing Model}

We have only an approximate measure of the lensing magnification from
the foreground clusters in the HFF clusters and therefore we could
expect uncertainties from the lensing models to impact our derived
size vs. luminosity results.  In particular, errors in the recovered
properties of lensed sources are such as to scatter sources so that
their measured half-light radii $r$ would show a steeper relationship
vs. luminosity $L$.  For cases where errors in the lensing model
impacted either both magnification axes equally or just a single
magnification axis, sources would scatter such that $r\propto L^{0.5}$
or $r\propto L$, respectively.

One way of illustrating the impact these uncertainties would have on
our results is to create a mock set of sources behind the first four
HFF clusters based on the Shibuya et al.\ (2015) size-luminosity
relation, the Bouwens et al.\ (2017b) $z\sim6$ $UV$ LF, and an assumed
scatter around the size-luminosity relation of 0.22 dex (as Shibuya et
al.\ 2015 find).  In creating the mock data set with apparent
magnitudes and sizes for individual sources, we will assume the CATS
v4.1 lensing models represent the truth and then interpret the results
using a median of the other parametric lensing models.  This mirrors
the forward-modeling approach we previously utilized in Bouwens et
al.\ (2017b) to derive $z\sim6$ LF results from our catalogs of
$z\sim6$ sources behind the first four HFF clusters.

Figure~\ref{fig:lenserror} illustrates the impact of these uncertainties
on the size-luminosity relation.  The left panel in this figure shows
the input distribution of sizes and luminosities, while the right
panel shows the recovered distribution after using a median model to
interpret the mock data set.  The red solid squares show the median
half-light radius recovered per 1-mag bin of $UV$ luminosity.

\begin{figure*}
\epsscale{1.17}
\plotone{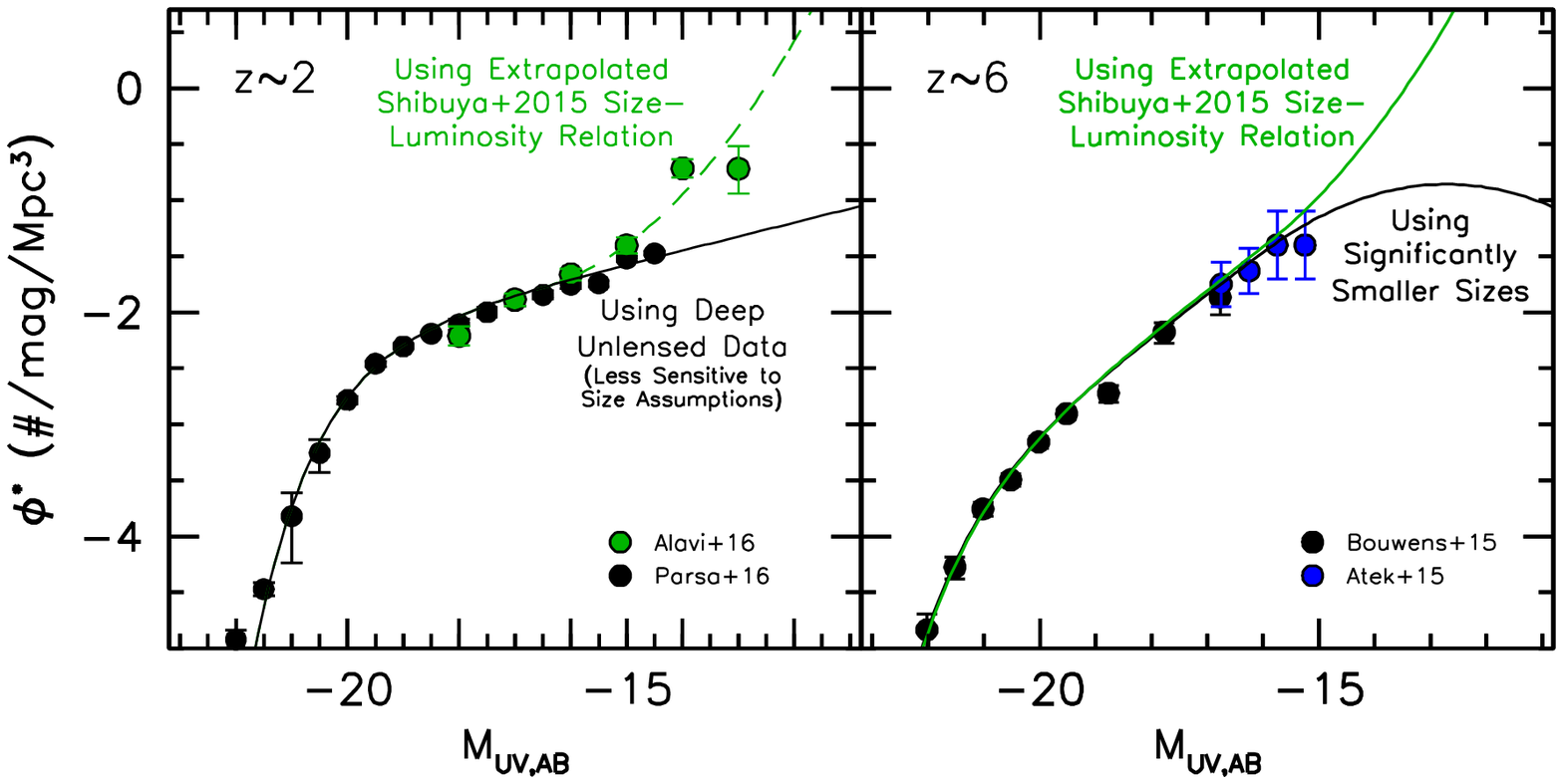}
\caption{An illustration of the significant impact the galaxy size
  distribution has on the faint-end ($>$$-$15 mag) form of the galaxy
  $z=2$-8 LFs.  Given that the requisite completeness corrections for
  LF determinations are directly calculable from the assumed
  size-luminosity relation, presumptions regarding the faint-end form
  of the LF are directly connected to what supposes the size
  distribution of faint galaxies to be.  If faint $z\sim2$ and
  $z\sim6$ galaxies have sizes which are a simple extrapolation of the
  Shibuya et al.\ (2015) size-luminosity relation, the recovered $UV$
  LFs at $z\sim2$ and $z\sim6$ combining blank field and lensing
  cluster observations are as indicated by the green lines and points
  (Alavi et al.\ 2016; \S5.4 of Bouwens et al.\ 2017b).  Meanwhile, if
  faint galaxies are assumed to have significantly smaller sizes than
  inferred from an extrapolation of the Shibuya et al.\ (2015) -- or
  equivalently a break in the size-luminosity relation: see magenta
  dashed line in the lower left panel of Figure~\ref{fig:msre0} -- the
  recovered $UV$ LFs show much lower volume densities.  The black line
  in the right panel are the Bouwens et al.\ (2017b) LF results and
  rely on significantly smaller size assumptions than the extrapolated
  Shibuya et al.\ (2015) relation.  The right panel also shows the
  blank field $z\sim6$ LF results from Bouwens et al.\ (2015) along
  with the results of Atek et al.\ (2015).  Meanwhile, the black line
  in the left panel are from Parsa et al.\ (2016) derive from the
  sensitive blank field observations over the HUDF (\textit{black line
    and black circles}) where size assumptions are not especially
  important at the faint end where sources are smaller than the PSF.
  If we suppose -- following most theoretical models -- that the $UV$
  LF at $z\sim2$ and $z\sim6$ extends towards fainter luminosities
  with a fixed (or progressively flatter) faint end slope, then the
  size-luminosity relation cannot extend to the lowest luminosity
  galaxies following the Shibuya et al.\ (2015) scaling, but must show
  a break at some luminosity towards a steeper
  scaling.\label{fig:lfshape}}
\end{figure*}

Comparisons of the left and right panels from
Figure~\ref{fig:lenserror} show the fairly dramatic impact of the
lensing model uncertainties on the recovered sizes or luminosities for
specific sources.  Brightward of $-15$ mag, the median recovered size
in a luminosity bin is very close to that from the input model.
However, faintward of $-$15 mag, the median recovered sizes in a
luminosity bin become substantially smaller.  This is due to the fact
that at lower luminosities, many sources from bright magnitude bins
are both scattering into the lower luminosity bins and scattering
towards smaller sizes, such that the median sizes are substantially
smaller than the input sizes.  One can see how brighter sources can
contaminate the lowest-luminosity bins inspecting the results from
Appendix B -- where it is shown that the true median luminosity of
sources in the $-12.5$-mag, $-13.5$-mag, and $-14.5$-mag bins are all
$\sim$$-$15 mag.

In addition, the results from Appendix B allow us to assess the impact
of errors in the lensing models on the number of extremely small
sources.  From the input model, only $\sim$1 source in an input sample
of $\sim$300 sources would be expected to have a half-light radius
$<$40 pc.  However, after including uncertainties from the lensing
models, $\sim$15 sources are expected to have such small inferred
sizes, indicating that some of the nominally small sources in our
sample may be there due to observational scatter.  For comparison,
there are 26 such sources in our own sample with half-light radii
$<$40 pc.  Therefore, our recovered total using the actual
observations is larger than the simulated total by 2$\sigma$.  This
suggests that some of the nominally compact sources do indeed have
sizes $<$40 pc, but many may actually be somewhat larger.

\begin{figure*}
\epsscale{1.15}
\plotone{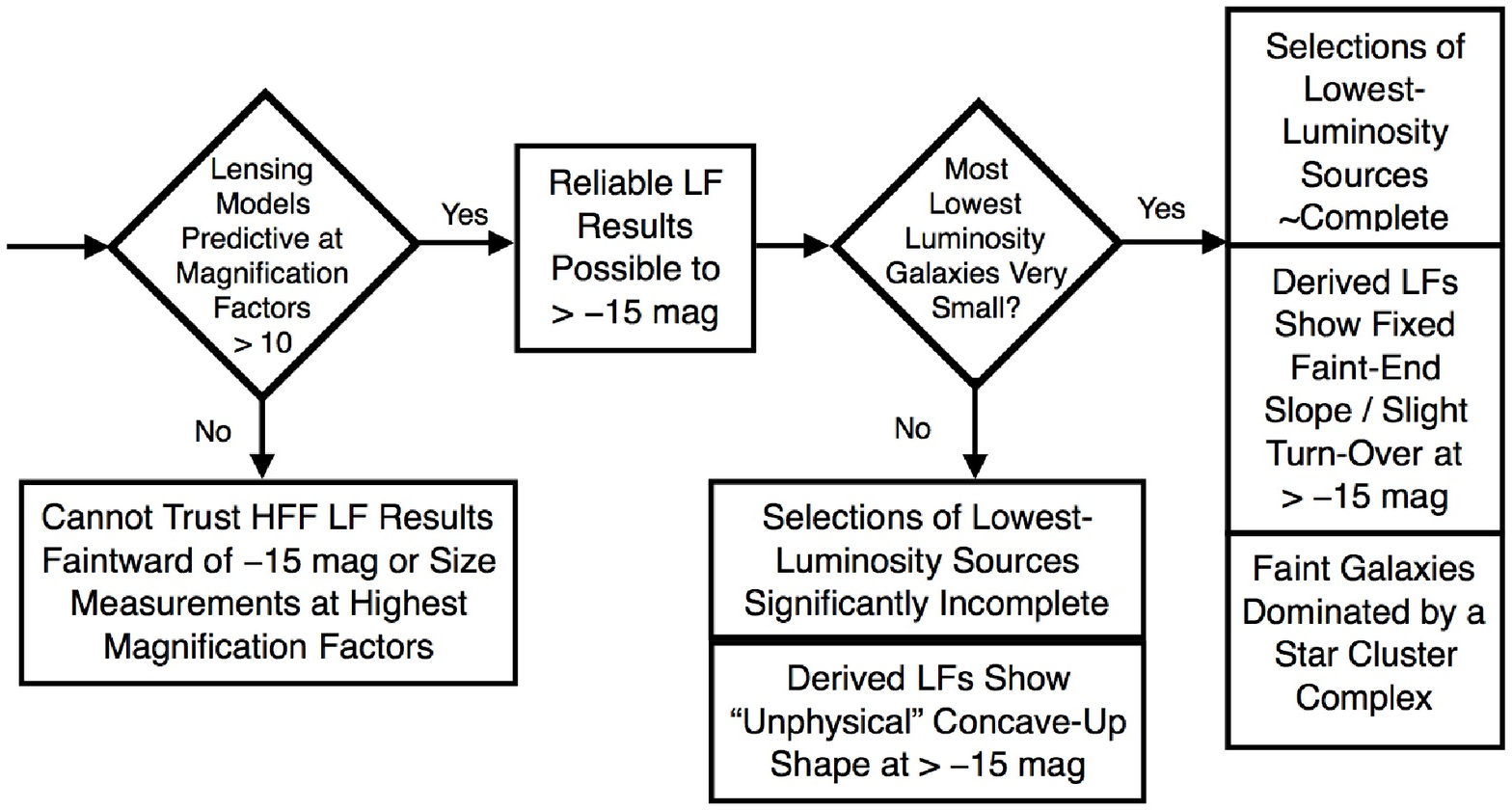}
\caption{A simple flowchart summarizing the connection between the
  form of the $UV$ LF at high-redshift and the implied size
  distribution for lower luminosity galaxies (see \S4.4).  As the
  various possibilities summarized by this logical flowchart rely on
  the HFF lensing models being predictive to magnification factors of
  $>$10 (e.g., as the Meneghetti et al.\ 2017; Prieuwe et al.\ 2017;
  Bouwens et al.\ 2017b results suggest), such a condition is included
  as the first step in the decision tree.  If we assume that the
  lower-luminosity galaxies have sizes that simply follow an
  extrapolation of the Shibuya et al.\ (2015) size-luminosity relation
  (where $r\propto L^{0.25}$), this implies a $UV$ LF with
  concave-upwards form at $>$$-$15 mag (see Figure~\ref{fig:lfshape}).
  On the other hand, if one supposes one should recover a standard
  faint-end form for the $UV$ LF, one must assume a steep
  size-luminosity relation, e.g., $r\propto L^{0.5}$.  The
  observations do not appear to allow for the assumption of both (1) a
  conventional size-luminosity relation (with $r\propto L^{0.25}$) and
  (2) a conventional faint-end form for the $UV$ LF at $>$$-$15
  mag.\label{fig:flowchart}}
\end{figure*}

The present exercise illustrates the need for caution in making claims
about specific small star-forming sources behind lensing clusters.

\subsection{Implications from the Faint End Form of the $z\sim6$ LF Derived from the HFFs}

There is a direct connection between (1) the distribution of sizes and
surface brightnesses assumed for the lowest luminosity galaxies and
(2) the faint-end form inferred for the $UV$ LFs at $z\sim2$-6 (see
Figures~\ref{fig:lfshape} and \ref{fig:flowchart}).  The purpose of
this section is to spell out this connection and the impact one has
for the other.

\subsubsection{Implications of Standard Shallow Size-Luminosity Relations for the Faint-End Form of the $UV$ LFs}

As we previously discussed in \S3.1 above, blank-field studies have
found the median half-light radius of brighter galaxies depends on the
luminosity $L$ of galaxies as $R\propto L^{0.25}$ (Shibuya et
al.\ 2015), across a wide range of redshifts.  Huang et al.\ (2013)
find a similar scaling at $z\sim4$ and $z\sim5$, and we might expect
similar scalings to apply to higher redshift galaxies if we
extrapolate the size-mass relations obtained by van der Wel et
al.\ (2014).  The well-known Kravtsov (2013) relation between halo
mass and galaxy size (their Figure 1) also argues for approximately
such a scaling.  Finally, both theoretical models (e.g., Liu et
al.\ 2017) and also some high-resolution hydrodynamical simulations,
e.g., Ma et al.\ (2017), report recovering almost exactly this scaling
in star-forming galaxies at high redshifts to very low luminosities.

If these scalings apply to extremely low luminosity $z\sim6$-8
galaxies, the surface brightness should vary as $L/R^2 \propto
L^{0.5}$.  With such a scaling, 0.001$L^*$ ($-$13.5 mag) galaxies
would have surface brightnesses 30$\times$ lower than $L^*$ galaxies
have.  At such low surface brightnesses, we would expect searches for
faint $z=2$-8 galaxies to be highly incomplete.  This would translate
into significantly lower surface densities of $z\sim6$ candidates in
the highest magnification regions, relative to that seen in lower
magnification regions.

Are such a deficit of sources seen in the very high magnification
($\mu>10$) regions relative to lower magnification ($\mu < 5$)
regions?  In Bouwens et al.\ (2017b), we find essentially an identical
surface density of $z\sim6$ sources in both low and high magnification
regions.  Ishigaki et al.\ (2017) also find a high surface density of
$z\sim6$ galaxies to $\sim$29 mag in their catalogs even in high
magnification $\mu>18$ regions, i.e., their Figure 1.

If we apply the expected high incompleteness in high magnification
regions (from the extrapolated Shibuya et al.\ 2015 relation) to the
Bouwens et al.\ (2017b) search results, we would infer very high
volume densities for the ultra-low luminosity sources at $z\sim6$.
In fact, this would translate into a concave-upwards faint-end form
for the $UV$ LF at $z\sim6$, as was inferred in \S5.4 of Bouwens et
al.\ (2017b) also applying the extrapolated Shibuya et al.\ (2015)
size-luminosity relation.  This is illustrated with the green solid
line in the right panel of Figure~\ref{fig:lfshape}.

Earlier, applying an extrapolation of the size-luminosity relation
obtained by Shibuya et al.\ (2015) for $z\sim2$ galaxies -- with a
similar size-luminosity dependence to their $z\sim6$ results -- Alavi
et al.\ (2016) had derived a $UV$ LF at $z\sim2$ showing exactly such
a concave-upward form.  This is indicated with the green solid line in
the left panel of Figure~\ref{fig:lfshape}.

Similar to the large-size analysis provided by the Bouwens et
al.\ (2017b) in their \S5.4, Atek et al.\ (2015) and Castellano et
al.\ (2016) made use of standard shallow size-luminosity relations in
deriving LF at $z\sim6$-7, only obtaining plausible LF results through
the restriction of their determinations to sources brightward of $-$15
mag.  H. Atek (private communication) indicated to us that they did
not extend their LF results faintward of $-$15 mag, due to
uncertainties in extrapolating the size-luminosity relation into this
regime and the very high volume densities implied at such faint
magnitudes by the uncertain incompleteness corrections.

While Atek et al. (2015) did not discuss the prevalence of $>$$−$15
mag galaxies at $z\sim6$-8, Livermore et al. (2017) show results down
to $-$12.5 mag.  The sizes that Livermore et al.\ (2017) quote for
their faint galaxies correspond to a median size of 0.5 kpc for their
faint sample, but as noted previously (Bouwens et al. 2017a,b)
incompleteness effects would be extreme for such large sizes.  The
surface brightness of $-$12.5-mag (0.0004$L^*$) sources would be
2500$\times$ lower than for $L^*$ sources and result in very high
completeness corrections at the low luminosity end.  When we carry out
a comparable analysis to Livermore et al.\ (2017) we find that we
cannot reproduce their derived luminosity function with such large
sizes (see also Kawamata et al.\ 2017); we can only broadly reproduce
their results when we use a smaller size distribution.  It is not
clear what the reason is for this discrepancy, but we note that the
combination of the median size and derived LF in Livermore et
al.\ (2017) do not fit comfortably in the ``flow chart'' of
Figure~\ref{fig:flowchart} here.

\subsubsection{Possibility of a Steep Size-Luminosity Relation?}

While one would expect to derive a ``concave-upwards'' luminosity
function for galaxies at $z\sim6$ making use of the standard shallow
size-luminosity relation for completeness measures, there are strong
observational and theoretical reasons for disfavoring such a
``concave-upwards'' luminosity function.  As demonstrated by Weisz et
al.\ (2014) and Boylan-Kolchin et al.\ (2014, 2015), abundance
matching of nearby dwarf galaxies sets a strong upper limit on the
volume density of lower luminosity galaxies in the high-redshift
universe.

From a theoretical perspective, one would expect the faint-end of the
LF to trace the halo LF to some degree, but at the extreme faint end,
the $UV$ LF is expected to flatten or even turn over, as a result of
increasingly inefficient gas cooling and radiative heating.  A typical
turn-over luminosity is $\sim-12$ mag (Liu et al.\ 2017; Finlator et
al.\ 2015; Gnedin 2016; O'Shea et al.\ 2015; Ocvirk et al.\ 2016; Yue
et al.\ 2016; Dayal et al.\ 2014).  Theoretical LFs are not expected
to become steeper towards the extreme low luminosity end.

If we discount such an upward change in the slope on the basis of
these plausibility arguments, we must assume that the size-luminosity
relation must show a break at $\sim$$-$17 mag, such that lower
luminosity galaxies are all very small.  This would translate to
generally high levels of completeness in searches for lower luminosity
galaxies.  Bouwens et al.\ (2017b) made use of such small size
assumptions in deriving constraints on $z\sim6$ $UV$ LF, finding a
roughly fixed faint-end slope to very low luminosities $>$$-$14 mag,
with a possible turn-over at the faint end.  The best-fit $z\sim6$ LF
results of Bouwens et al.\ (2017b) are included in the right panel of
Figure~\ref{fig:lfshape} with a black line.

In Bouwens et al.\ (2017a), we had provided an independent motivation
for supposing that faint galaxy population is intrinsically small --
and the observed size-luminosity relation is not predominantly driven
by surface-brightness selection effects.  That motivation is the
approximately constant surface density of $z\sim6$-8 sources in
high-magnification regions behind clusters over a wide range of shear
factors.  This would only be the case if sources were intrinsically
small, as larger intrinsic sizes would result in a much higher
prevalence of sources in high-magnification regions with low shear
factors.  This is due to the fact that sources are readily detectable
in low shear regions over a much larger range of sizes than is
possible in regions around a lensing cluster with high shear.

As a caveat to this discussion, we should emphasize that the
conclusions that we have drawn in this subsection are sensitive to the
predictive power of the lensing models.  If the lensing models lose
their predictive power above magnification factors of $\sim$10, the
sources that make up our nominally lowest luminosity samples (i.e.,
$M_{UV}>-15$ mag or $M_{UV}>-14$ mag) would instead be prominently
made up of sources at higher intrinsic luminosities, i.e.,
$M_{UV}\sim-15$ mag, scattering to lower lower luminosities due to
uncertainties in the lensing models.  Despite this possibility, we
emphasize that there is significant evidence that lensing models
(especially the median model) maintain their predictive power to
magnification factors of at least 20-30, if the tests run by
Meneghetti et al.\ (2017), Prieuwe et al.\ (2017), or Bouwens et
al.\ (2017b) can be trusted.

The arguments presented in this section are subtle but clear and are
summarized in Figure~\ref{fig:flowchart}. If plausible luminosity
functions are to be obtained at extremely low luminosities (i.e.,
$>$$-$15 mag), where plausible means ``not upturning'' (as predicted
in most theoretical models: e.g., Dayal et al.\ 2014; Gnedin 2016; Liu
et al.\ 2016), then such sources must be small.  The use of sizes
resulting from a simple extrapolation of the size-luminosity relation
found for higher luminosity galaxies would suggest very large
completeness corrections and imply an upturn.  The clear implication
is that there must be a break in the size-luminosity relation below
$\sim-17$ mag to a steeper slope at lower luminosities.

\section{Discussion}

\subsection{Comparison with Previous Compilations of Small Star-Forming Sources}

Before discussing the implications of the recovered size distribution
for the sample of $z=6$-8 galaxies we have identified, it is useful to
reexamine the sample of very small ($\sim$10-100 pc) sources
identified by Vanzella et al.\ (2017a) using the current size
measurements and also compare against new results recently obtained by
Kawamata et al.\ (2017: which are an update to the earlier Kawamata et
al.\ 2015 results).

Encouragingly, two of the three star-forming candidates that Vanzella
et al.\ (2017a) identify over MACS0416 (the only HFF cluster analyzed
both in the present study and that earlier study) are also included in
our compilation of small sources (Table~\ref{tab:tiny}).
M0416I-6115434445 corresponds to GC1 from Vanzella et al.\ (2017a),
while M0416I-6114803434 corresponds to D1 from Vanzella et
al.\ (2017a).  We also have D2 from Vanzella et al.\ (2017a) in our
source catalogs and so we can also compare our size and luminosity
measurements for these sources.

For GC1, D1, and D2, we infer half-light radii of 21$_{-13}^{+31}$ pc,
38$_{-14}^{+21}$ pc, and 72$_{-30}^{+39}$ pc vs. similar half-light
radius measurements of 16$\pm$7 pc, 140$\pm$13 pc, and $<$100 pc,
respectively, from Vanzella et al.\ (2017a).  The sizes we infer for
M0416I-6115434445/GC1 and M0416I-6103003258/D2 are in excellent
agreement with those from Vanzella et al.\ (2017a).  For the third
source M0416I-6114803434, the half-light radius we estimate is
$3\times$ higher than the estimate from Vanzella et al.\ (2017a).
This is almost certainly due to their fit giving a best-fit Sersic
parameter of $n=3.0\pm0.3$ (which results in a much larger half-light
radius estimate).  However, Vanzella et al.\ (2017a) do quote a 50-pc
estimated size for the central high-surface brightness region of that
source.

\begin{figure}
\epsscale{1.18}
\plotone{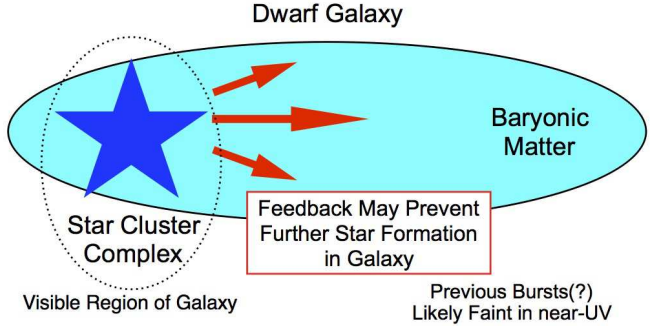}
\caption{Cartoon schematic showing the formation of a single star
  cluster complex within a high-redshift dwarf galaxy.  The star
  cluster complex forms out of overdense baryonic material.  As
  feedback from the star cluster complex could temporarily inhibit
  star formation in other regions of the galaxy (e.g., Bastian 2008),
  galaxies may be much smaller in terms of their readily-visible
  spatial extent than they actually are.  The spatial size of distant
  star-forming galaxies would also be made to look smaller than they
  are due to dominant impact of the youngest star cluster complexes on
  the $UV$ morphologies of dwarf galaxies (e.g., Ma et al.\ 2017), as
  occurs e.g. in nearby tadpole galaxies (e.g. Kiso 5639: Elmegreen et
  al.\ 2016) or blue compact dwarf galaxies (Elmegreen et al.\ 2012b;
  Papaderos et al.\ 2008).  In particular, Figures 4 and 5 of
  Elmegreen et al.\ (2016) would be strikingly similar to the compact
  objects seen at high redshift.  The lensed $z\sim5$ galaxy MS1358+62
  (Franx et al.\ 1997), with a dominant star cluster complex $<$200 pc
  in size (see Swinbank et al.\ 2009; Zitrin et al.\ 2011), also
  provides us with another dramatic example.\label{fig:cartoon}}
\end{figure}

\begin{figure*}
\epsscale{1.1}
\plotone{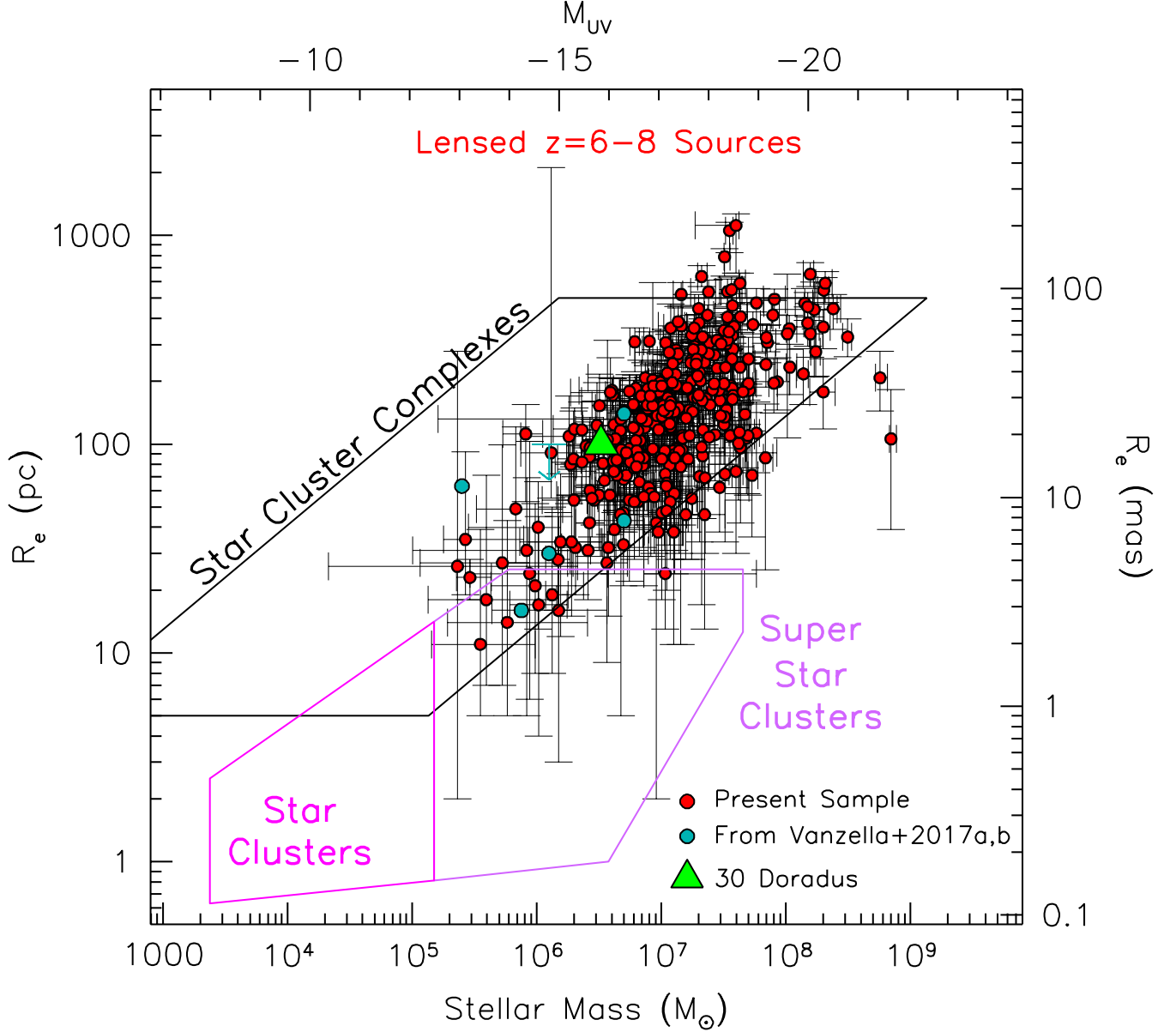}
\caption{Comparison of the inferred sizes and luminosities of lensed
  galaxies in the HFF clusters (\textit{red circles}) with star
  clusters (\textit{demarcated by the magenta lines:} \S3.3), super
  star clusters (\textit{demarcated by the violet lines:} \S3.3), and
  star cluster complexes (\textit{demarcated by the black lines:}
  \S3.2 and Figure~\ref{fig:msre00}).  $1\sigma$ errors are the same
  as shown in Figure~\ref{fig:msre0}.  The cyan circles show the sizes
  and luminosities reported by proto-globular clusters and star
  cluster candidates claimed by Vanzella et al.\ (2017a, 2017b) while
  the green triangle shows the size and luminosity of the 30 Doradus
  star complex.  The conversion between a given $UV$ luminosity and a
  stellar mass is made assuming a star formation duration of 10 Myr.
  While most of the lensed $z=6$-8 sources in the HFF observations
  appear to have sizes and luminosities consistent with star cluster
  complexes seen in $z=0$-3 galaxies (Bastian et al.\ 2006; Jones et
  al.\ 2010; Wisnioski et al.\ 2012; Swinbank et al.\ 2012; Livermore
  et al.\ 2012; Adamo et al.\ 2013; Livermore et al.\ 2015; Johnson et
  al.\ 2017; Dessauges-Zavadsky et al.\ 2017), a few source lie in the
  star cluster regions.  We use the few sources found in the star
  cluster region to place constraints on the volume density of
  proto-globular clusters at $z\sim6$ (\S5.4 and
  Figure~\ref{fig:gclf}).\label{fig:msre_sc}}
\end{figure*}

We also compare our size measurements with those from Kawamata et
al.\ (2017), who have updated the results from Kawamata et al.\ (2015)
to include sources from all six HFF clusters and parallel fields.
Cross-matching our source catalogs with sources in the Ishigaki et
al.\ (2017)/Kawamata et al.\ (2017) catalogs, we find 80 sources in
common.  In the median, our size measurements agree fairly well with
those from Kawamata et al.\ (2017), with our measured sizes being
15$\pm$7\% larger.  For individual sources, the differences are
larger, with a $1\sigma$ scatter in our size estimates of 0.32 dex.
For sources in common between our catalogs, when we estimate sizes
less than 50 pc, the median size measurement in their catalog is 35
pc.  Similarly, when Kawamata et al.\ (2017) estimate sizes less than
50 pc, the median size measurement in our catalog is 62 pc.  As such,
there is reasonable agreement (at least in the median) between our
estimated sizes and those of Kawamata et al.\ (2017) and also our
selected samples of sources with small sizes and those of Kawamata et
al.\ (2017).  This is encouraging and increases our confidence in our
results as we proceed to an interpretation.

\subsection{Similarity of Lensed $z=6$-8 Sources to Star Cluster Complexes in $z\sim1$-3 Galaxies}

\begin{figure*}
\epsscale{1.1}
\plotone{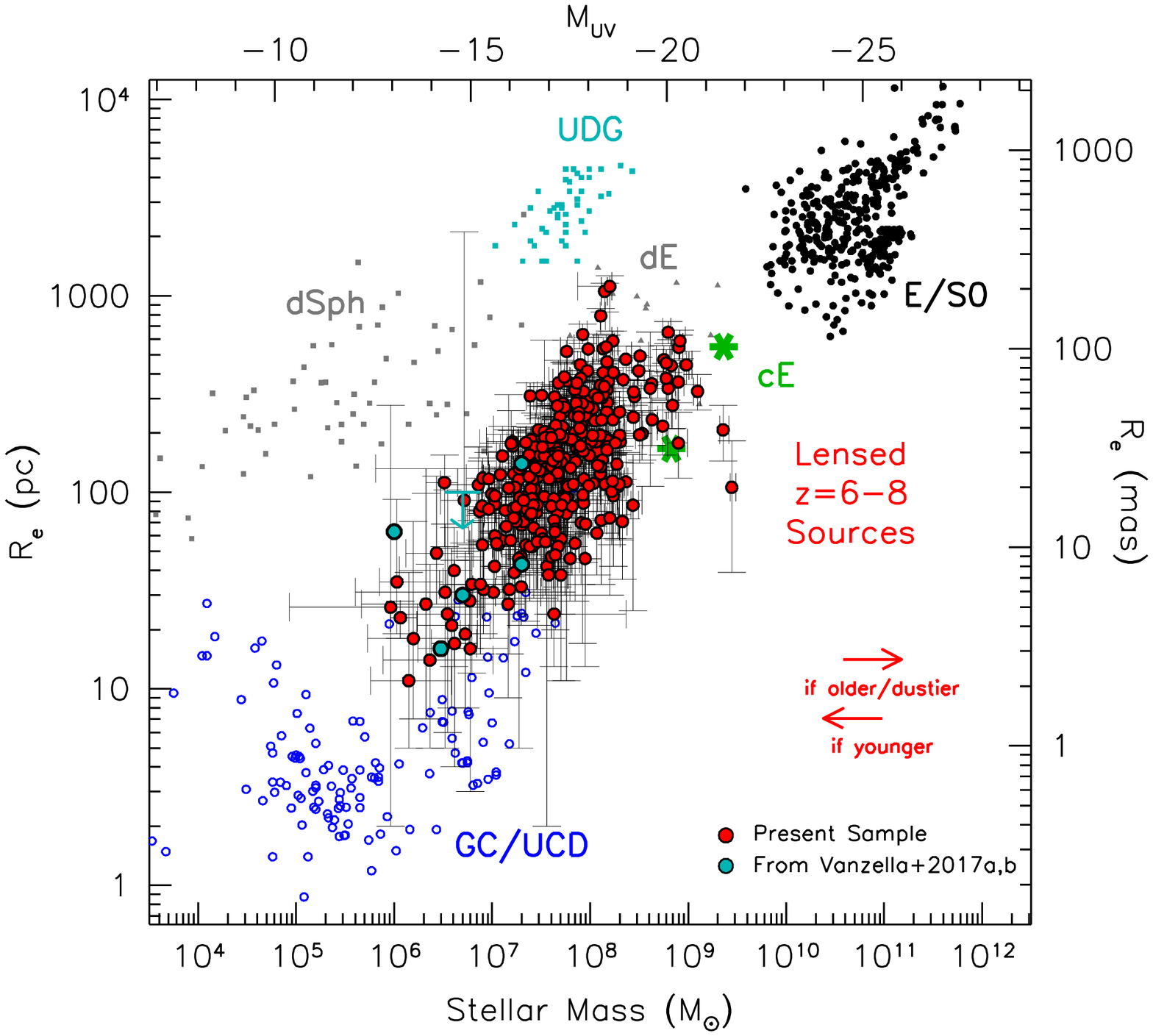}
\caption{Similar to Figure~\ref{fig:msre_sc}, but focusing on
  comparisons with the sizes and masses of various evolved stellar
  systems in the nearby universe, including E/S0 (\textit{black
    circles}), ultra diffuse elliptical galaxies (\textit{cyan
    squares}), dwarf spheroids (dSphs: \textit{gray squares}), dwarf
  ellipticals (dEs: \textit{gray triangles}), compact ellipticals
  (cEs: \textit{green star-like symbols}), and globular clusters/ultra
  compact dwarfs (\textit{open blue circles}).  The conversion between
  a given $UV$ luminosity and a stellar mass is made assuming a star
  formation duration of 100 Myr.  The left and right red arrows shows
  the expected change in the inferred masses when changing the
  duration of star formation from 100 Myr to 10 Myr and 400 Myr,
  respectively.  The layout of this figure is similar to Figure 8 in
  Brodie et al.\ (2011) and Figures 11 of Norris et al.\ (2014).
  While the smallest lensed $z=6$-8 sources in the HFF observations
  appear to have sizes / luminosities consistent with that of globular
  clusters / ultra-compact dwarfs, most of the lensed $z=6$-8 galaxies
  have sizes and luminosities that lie in the region between globular
  clusters and that of elliptical galaxies and seem to best match that
  seen in star cluster complexes (Figure~\ref{fig:msre_sc}).  They
  presumably undergo further dynamical evolution and/or accretion
  before becoming the evolved descendants we see
  today.\label{fig:msre}}
\end{figure*}

As we discussed in \S4.2 to \S4.4, the ultra faint sources we identify
at $z\sim6$-8 behind the HFF clusters show significantly smaller sizes
than the extrapolated size-luminosity relation for $z=6$-8 galaxies
from blank field studies.  Interestingly, the size and luminosities of
these sources lie in the general range of star cluster complexes
identified in $z\sim2$-3 galaxies, as presented earlier in
Figure~\ref{fig:msre00} in \S3.2 (see also Figure~\ref{fig:msre0}).
It is therefore logical to wonder if some of the sources we are
identifying behind the HFF clusters may simply be star cluster
complexes viewed at $z\sim6$-8.

In some cases, it is possible that these sources correspond not simply
to individual star cluster complexes but actually to super star
clusters.  It is difficult to be sure about a star cluster
identification at the spatial resolutions available with HST (or
ground-based telescopes) even with lensing magnification.
Nevertheless, the small sizes, young ages, and almost identical
redshifts to brighter nearby galaxies does make such an identification
at least possible, as Vanzella et al.\ (2017b) do for several compact
star-forming sources in the MACS0416 field.

Whatever the reality be for the smallest sources in our sample, it
certainly seems plausible to make the connection of sources in our
study to individual star cluster complexes.  The shear number of
sources with $\sim$100 pc sizes and luminosities similar to star
cluster complexes makes the connection natural.  The fact that the
observed sizes of the sources are smaller than the relation seen for
the brightest galaxies and from lower redshifts -- where radius scales
as $L^{1/4}$ or $L^{1/3}$ (e.g., de Jong \& Lacey 2000; van der Wel et
al.\ 2014) suggests we may not be seeing all the baryonic material
associated with a given dwarf galaxy.  Indeed, we are likely only
observing a single dominant star cluster complex within each source.
Consistent with our suggested scenario, observations show an
increasing fraction of the light in star cluster complexes, from low
redshift to high redshift (Ribeiro et al.\ 2016).

How likely is it for lower-mass galaxies in the $z=6$-8 universe to
host just a single dominant star cluster complex?  While addressing
such a question would almost certainly require high resolution
hydrodynamical simulations (e.g., see Ma et al.\ 2017), one could
easily imagine the collapse of an overdensity resulting in the
formation of a star cluster complex and feedback from that star
cluster complex preventing star formation from occurring at any other
position in a dwarf galaxy (Figure~\ref{fig:cartoon}).  A simple
calculation assuming SNe wind speeds of 50$\,$km/s and a dwarf galaxy
size of 200 pc suggests a feedback time of only 4 Myr, potentially a
short enough time for starburst activity in one star cluster complex
to regulate star formation across an entire dwarf galaxy (e.g., see
Bastian 2008).

The discussion here and the match with local and lower-redshift star
cluster complexes (discussed in the next section and shown in
Figure~\ref{fig:msre_sc}) suggest that low-luminosity high-redshift
$z\sim6$-8 galaxies might be well-described as having a single
dominant star cluster complex.  Such a star forming cluster complex
does not, of course, preclude the galaxy itself from being larger
(cf., the $z=4.92$ Franx et al.\ (2017) MS1358+62 example) in terms of
its physical extent, but the lower surface brightness regions could
easily be missed in many sources due to cosmic surface brightness
dimming (e.g., see Ma et al.\ 2017).

\subsection{Connection to Sources in the Local Universe}

Based on the considerations from the previous section, it is already
clear that some of the faint lensed sources in the HFFs could
correspond to forming star cluster complexes in the distant universe,
whether those star cluster complexes are the dominant (and only?)
complex in a galaxy or whether those complexes are associated with a
brighter system.

It is interesting to ask how the faint lensed sources we are finding
compare with various stellar systems found in the nearby universe.
For this exercise, we use the compilation that Norris et al.\ (2014)
and M. Norris (2017, private communication) provide of the sizes and
masses for a wide variety of local sources.  This compilation includes
elliptical galaxies (e.g., Cappellari et al.\ 2011; McDermid et
al.\ 2015), ultra-diffuse ellliptical galaxies (e.g., van Dokkum et
al.\ 2015), dwarf ellipticals and spheroids (e.g., Misgeld et
al.\ 2008), compact ellipticals such as Messier 32 (e.g., Chilingarian
et al.\ 2009), ultra-compact dwarfs (e.g., Evstigneeva et al.\ 2007;
Misgeld et al.\ 2011), and globular clusters (e.g., Hasegan et
al.\ 2005; Firth et al.\ 2007; Mieske et al.\ 2007; Francis et
al.\ 2012).  The black and magenta lines indicate the region in
parameter space where we would expect star cluster complexes and star
clusters, respectively, to reside (\S3.2-\S3.3).

Figure~\ref{fig:msre} shows the inferred sizes and indicate masses for
our lensed $z\sim6$-8 sample relative to the Norris et al.\ (2014)
compilation.  The indicative masses that we use for our lensed sample
are computed assuming a fixed stellar population duration of 100 Myr
in converting from their inferred $UV$ luminosities $M_{UV}$.

Interestingly, some lensed sources in our samples have size and
luminosities in the regime of ultra-compact dwarf galaxies or globular
clusters, with measured sizes $<$40 pc -- as we previously remarked in
\S4.2-\S4.4 and suggestively indicated by the magenta lines.  Kawamata
et al.\ (2015) had previously reported two sources with such small
sizes.  Now, thanks mostly to the present work and that by Kawamata et
al.\ (2017) as well as a few candidates by Vanzella et al.\ (2017a,
2017b), we now know of a large number of very small star-forming
sources in the distant universe.

While we must allow for the fact that some fraction of these
high-redshift sources might be intrinsically larger than what we infer
due to lensing uncertainties (see \S4.4), it is likely that a modest
fraction of these sources may genuinely be quite small.  What
therefore is the nature of these especially compact star-forming
sources?  A few could, in fact, correspond to proto-globular clusters
or super star clusters, as Figure~\ref{fig:msre_sc} illustrates (see
also Vanzella et al.\ 2017a, 2017b).  Nevertheless, we should
emphasize that the bulk of our sample is more extended in size,
suggesting that a more natural hypothesis is that most sources better
match up with the properties of star cluster complexes.

\begin{deluxetable}{cc}
\tablecolumns{2}
\tabletypesize{\footnotesize}
\tablecaption{Volume Density Constraints on proto-globular clusters forming at $z\sim6$\label{tab:gclf}}
\tablehead{\colhead{$M_{UV}$} & \colhead{$\phi$ (Mpc$^{-3}$)}}
\startdata
\multicolumn{2}{c}{This Work\tablenotemark{a}}\\
$-$17.50 & $<$0.00013\tablenotemark{b}\\
$-$16.50 & $<$0.00033\tablenotemark{b}\\
$-$15.50 & $<$0.013\tablenotemark{b}\\
$-$14.50 & $<$0.065\tablenotemark{b}\\
$-$13.50 & $<$0.068\tablenotemark{b}\\
\\
\multicolumn{2}{c}{Estimated From Vanzella et al.\ (2017a)}\\
$-$15.50 & $<$0.0106\tablenotemark{b,c}\\
\\
\multicolumn{2}{c}{Volume Densities Probed with the HFF program\tablenotemark{d}}\\
$-$17.50 & 0.00014\\
$-$16.50 & 0.00041\\
$-$15.50 & 0.0019\\
$-$14.50 & 0.0098\\
$-$13.50 & 0.054\\
$-$12.50 & 0.8
\enddata
\tablenotetext{a}{Fraction of sources which are measured to have a size of 40 pc or smaller multiplied by the volume density of sources in the Bouwens et al.\ (2017b) $z\sim6$ LF}
\tablenotetext{b}{$1\sigma$ upper limits}
\tablenotetext{c}{We use the same criteria in establishing the upper limits on the proto-globular cluster volume densities as we use for our own observational results}
\tablenotetext{d}{Gray region in Figure~\ref{fig:gclf}}
\end{deluxetable}

\subsection{Limits on the Volume Densities of Forming Globular Clusters in the $z\sim6$ Universe}

We can also use our size measurements of lensed sources behind the HFF
clusters to set constraints on the luminosity function of
proto-globular clusters in the early universe.  As we remarked in the
introduction, the large ages of stars in most globular clusters
together with the high gas densities appropriate for globular cluster
formation (Goddard et al.\ 2010; Adamo et al.\ 2011; Silva-Villa \&
Larsen 2011) -- as well as the high prevalence of globular clusters
even in lower-mass galaxy halos (Spitler \& Forbes 2009; Harris et
al.\ 2013) -- strongly suggest a $z\gtrsim 1.5$ formation era.  Having
observational constraints on the formation of these sources in the
early universe is both valuable and interesting.

Given the proximity in time of powerful facilities like the James Webb
Space Telescope JWST, there are now numerous predictions for the
number of such clusters which might be found in a typical search field
with the JWST (Renzini 2017; Boylan-Kolchin 2017a, 2017b; Elmegreen et
al.\ 2012a) as well as candidate proto-globular clusters identified in
separate studies (Vanzella et al.\ 2017a).

\begin{figure}
\epsscale{1.15}
\plotone{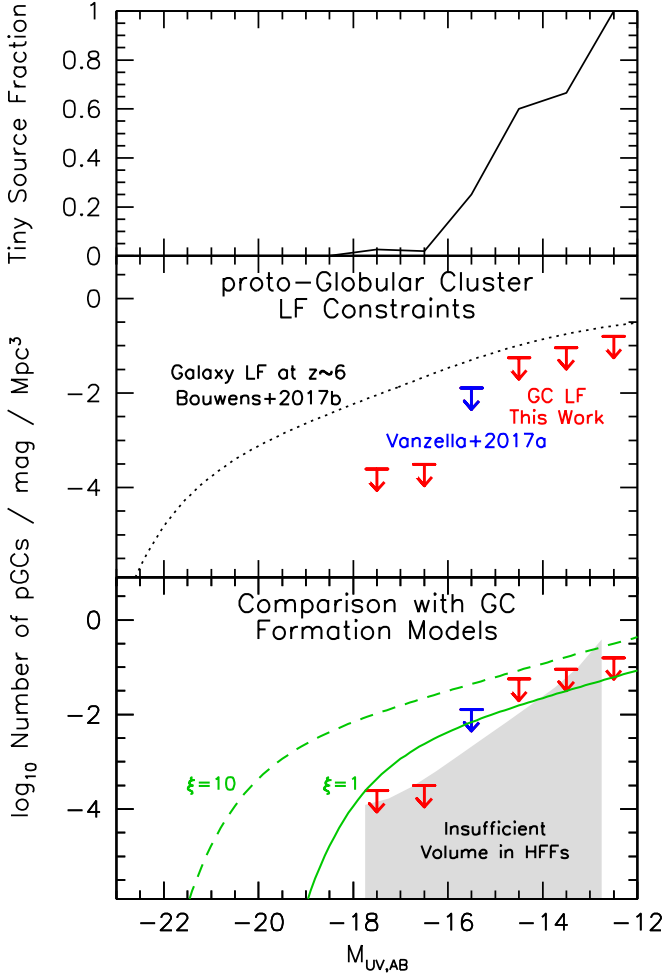}
\caption{(\textit{upper}) Fraction of the lensed $z\sim6$ sources in
  the HFF observations which are measured to have half-light radii
  $\leq$40 pc.  (\textit{middle}) Constraints on the volume density of
  forming proto-globular clusters at $z\sim6$ using searches for small
  sources behind the HFF clusters.  The plotted red upper limits
  combine the $UV$ LF constraints we obtained from the first four HFF
  clusters with the fraction of sources for which the available
  constraints suggest sizes $\leq$40 pc.  The blue upper limit give
  the volume density constraints we infer here for the proto-globular
  cluster reported by Vanzella et al.\ (2017a).  The dotted black line
  shows a recent determination of the $z\sim6$ LF for galaxies from
  the Hubble Frontier Fields from Bouwens et al.\ (2017b) and which
  was used to derive the observational constraints (together with the
  results in the top panel).  (\textit{lower}) Comparison of the
  observational constraints on the proto globular cluster volume
  densities with the predicted LF of proto-globular cluster candidates
  from Boylan-Kolchin (2017a, 2017b) assuming the mass ratio $\xi = 1$
  (\textit{green solid line}: see \S5.4) and $\xi = 10$ (\textit{green
    dashed line}).  The shaded gray region shows the volume densities
  and luminosities where the HFF program does not provide us with
  sufficient volume to probe; its upper envelope is equal to the
  reciprocal of the total volume computed to be available over the
  first six HFF clusters in the Bouwens et al.\ (2017, in prep)
  analysis.\label{fig:gclf}}
\end{figure}

To provide constraints on the volume density of forming globular
clusters in early universe, we explicitly consider the size
constraints we have available for our full sample of $z\sim6$ sources
over the first four HFF clusters from this paper vs. that expected for
star clusters (Figure~\ref{fig:msre_sc}).  If the size measurements we
have for a source yield a half-light radius measurement of $<$40 pc,
we consider it as a possible globular cluster candidate.  We are, of
course, aware that 40 pc is larger than the upper size limit on star
clusters in the lower redshift universe, which is the range of 10-20
pc (\S3.3), but we are considering a more inclusive selection here to
ensure that both our selection and upper limits we set include the
broadest range of possible candidates.  Using this definition, one of
the $z\sim6$ sources with size measurements from Kawamata et
al.\ (2015), i.e., HFF1C-i10, would qualify as a proto-globular
cluster candidate.

In the upper panel of Figure~\ref{fig:gclf}, we present the fraction
of $z\sim6$ sources which could correspond to forming globular
clusters vs. $UV$ luminosity (i.e., those with $r_e \leq 40$ pc: see
Table~\ref{tab:tiny}).  We then show in the middle panel of
Figure~\ref{fig:gclf}, the implied upper limits we can obtain on the
volume density of proto-globular clusters forming in the distant
universe.  Given the challenges in being sure that any given system
corresponds to a proto-globular cluster (or the challenges in being
sure that specific lensed sources are in fact small: see \S4.4), we
include our constraints as upper limits.  In the same panel, we also
include the one proto-globular cluster candidate GC1 identified by
Vanzella et al.\ (2017a) at $z\sim6$ again as an upper limit on the
volume density.  For consistency with the candidates included from our
own study, we only include this candidate since its inferred size of
16$\pm$7 pc satisifes our criterion of a source size $<$40 pc.  In
Table~\ref{tab:gclf}, we provide the estimated upper limits on the
volume density of sources for ready comparison.

For comparison with our proto-globular cluster constraints, we also
include on Figure~\ref{fig:gclf} the predicted number of
proto-globular cluster and evolved globular cluster systems as a
function of $UV$ luminosity estimated to be present in the $z=6$-10
universe using the model of Boylan-Kolchin (2017a, 2017b).
Boylan-Kolchin (2017b) provide a convenient Schechter function
approximation for this model globular cluster LF which we include in
the figures.  Shown are the predicted globular cluster LF for two
different values for the ratio of mass in the initial globular cluster
at its birth and that present at $z=0$, i.e., $<m_{GC}
(\textrm{birth})>/<m_{GC} (z=0)>$.  Following Boylan-Kolchin (2017a,
2017b), we use the symbol $\xi$ to describe this ratio and present
both the $\xi=1$ case (where $\phi^*=4 \times 10^{-3}$ Mpc$^{-3}$,
$M^*=-16.9$, $\alpha=-1.7$) and $\xi=10$ case (where $\phi^*=4 \times
10^{-3}$ Mpc$^{-3}$, $M^*=-19.4$, $\alpha=-1.7$) in
Figure~\ref{fig:gclf}.  The $\xi=10$ case involves substantial mass
loss after the initial globular burst would favor very bright
proto-globular clusters in the early universe.  Such a scenario is
motivated e.g. by Schaerer \& Charbonnel (2011) based on chemical
complexity of the enrichment in globular clusters (see also D'Ercole
et al.\ 2008; Renzini et al.\ 2015).

For context, we also show on Figure~\ref{fig:gclf} the volume
densities to which we would be able to search for proto-globular
clusters of specific luminosities with the full HFF program.  The
volume densities are computed as in Bouwens et al.\ (2017b) and hence
would be for a probe of proto-globular clusters at $z\sim6$, but are
based on the results over all six HFF clusters (Bouwens et al.\ 2017,
in prep).  These volume densities are also compiled in
Table~\ref{tab:gclf} for convenience.  The search volume available for
proto-globular clusters in the $z=6$-10 universe is $\sim$3$\times$
larger.

Remarkably, the predictions of the $\xi=1$ Boylan-Kolchin (2017a,
2017b) model lie very close to the upper limits we can set on the
basis of existing HFF search results faintward of $-$16 mag.  They are
clearly in excess of our constraints over the HFFs brightward of $-$16
mag, predicting $\sim$3 and $\sim$10 sources at $-17.5$ mag and
$-16.5$ mag, respectively, within the volume of the HFF program.
Interestingly, at slightly fainter luminosities, i.e., $-$15.5 mag,
our observational results (\textit{red downward arrows}) are also much
more consistent with the minimal $\xi=1$ scenario sketched out by
Boylan-Kolchin (2017a: \textit{green solid line}).  If the
Boylan-Kolchin (2017a) $\xi=1$ scenario is correct, three $-15.5$-mag
sources identified with the HFF program are expected to correspond to
proto-globular clusters in formation.  While a small number, these
constitute $\sim$50\% of the tiny star-forming sources that we have
identified at those low luminosities.  These results indicate that we
observe plausible consistency between the $\xi=1$ model of
Boylan-Kolchin (2017a) and what we derive from our size and LF results.

Interestingly, and rather definitively, the $\xi=10$ model of
Boylan-Kolchin (2017a) exceeds the upper limits we can set from the
HFF observations at all luminosities.  As such, we can probably
already rule the $\xi=10$ model out (unless proto-globular clusters
are lost within the light from their host galaxy).  This adds to other
independent evidence against such models (Bastian \& Lardo 2015;
Kruijssen 2015; Webb \& Leigh 2015; Martocchia et al.\ 2017; Elmegreen
2017).

This suggests that observers may be on the brink of exploring the
formation of globular clusters in the distant universe with current
and especially using future observations with JWST (see also
discussion in Renzini 2017).  With JWST, not only will be able to
search for proto-globular cluster candidates much more efficiently,
but we will be able to characterize each candidate system in detail
using high S/N spectrocopy, probing the velocity dispersion (and thus
allowing for a measurement of the dynamical mass) as well as the
chemical maturity of such systems.

\section{Summary}

Here we make use of the unique depth and resolving power of the HFF
cluster observations to examine the sizes and luminosities of 153
$z\sim6$, 101 $z\sim7$, and 53 $z\sim8$ sources identified in the
early universe behind the first four HFF clusters (307 $z=6$-8
galaxies in total).  We restricted ourselves to an analysis of sources
behind the first four HFF clusters since those possess the most mature
lensing models leveraging substantial spectroscopic redshift
constraints and a substantial number of multiple image systems.

The depth of the HFF observations and the lensing from the massive
foreground clusters make it possible for us to measure the sizes for
$\sim-18$ mag and $\sim-15$ mag galaxies to a typical $1\sigma$
accuracy of $\sim$50 pc and $\sim$10 pc, respectively.  Achieving such
high accuracy on size measurements is crucial for distinguishing
between normal galaxies, star cluster complexes, star clusters, and
even proto-globular clusters forming in the early universe.

To obtain the most robust measurements on the sizes and luminosities
of sources, we make use a MCMC procedure to fit the available imaging
data for each source (\S4.1).  We also utilize the median
magnification and shear factors derived from six different varieties
of parametric lensing models CATS, Sharon/Johnson, GLAFIC, Zitrin-NFW,
Keeton, and Caminha.  The model profile is lensed according to the
median magnification and shear factor, convolved with the PSF, and
then compared with a stack of the available WFC3/IR data on each
source.

The measured sizes of lensed $z=6$-8 galaxies in our sample trend with
$UV$ luminosity $L$ approximately as $L^{0.5\pm0.1}$
(Figure~\ref{fig:msre0}: see \S4.2) at $>$$-$17 mag, reaching sizes of
11$_{-6}^{+28}$ pc.  This is shallower than the trend for more
luminous ($<$$-$18 mag) sources, i.e., $r\propto L^{0.26\pm0.03}$,
suggesting a break in the relation at $\sim$$-$17 mag.  While the
trend we recover at lower luminosity is similar to what one might
expect as a result of surface brightness selection effects (e.g.,
Bouwens et al.\ 2017a; Ma et al.\ 2017), our lowest luminosity samples
show no particular evidence for being especially incomplete.  If one
assumes the lowest luminosity sources have small sizes (and hence
minimal completeness corrections), one derives plausible LF results,
as we demonstrated in Bouwens et al.\ (2017b) for the $z\sim6$ LF.  If
one however adopts a much shallower size-luminosity relation -- as
found by Shibuya et al.\ (2015) to holds for the lower-redshift
star-forming galaxies and the brightest high-redshift sources --
high-redshift selections would become appreciably incomplete at
$>$$-$15 mag and the inferred LFs would show a concave-upwards form at
$>$$-$15 mag (see \S4.4).  This situation is summarized in
Figures~\ref{fig:lfshape} and \ref{fig:flowchart}.

Sources in our lensed $z=6$-8 samples have measured sizes and
luminosities which are very similar to that derived for star cluster
complexes identified in galaxies at $z=0$-3 (Jones et al.\ 2010;
Livermore et al.\ 2012, 2015; Wisnioski et al.\ 2012; Swinbank et
al.\ 2012; Johnson et al.\ 2017).  In fact, the typical $-$15 mag
galaxy in our samples has a smaller half-light radius than 30 Doradus,
which has a measured half-light radius of $\sim$100 pc.  This could be
interpreted to suggest that lower luminosity galaxies in the early
universe may often contain a single prominent star cluster complexes
which dominates the observed UV morphology.

The impact that errors in the gravitational lensing models would have
on our results is considered using sophisticated simulations (\S4.3)
similar to that used in our previous work on the $z\sim6$ LF.  The
results of the simulations suggest that some fraction of the compact
star-forming sources identified in our fields actually have larger
physical sizes and are simply inferred to be small, due to errors in
the estimated lensing magnification.  Results from these simulations
also show fewer compact sources than we recover in the actual
observations, suggesting that a fraction of the small sources we
identify are bona-fide.

We also place the measured size and luminosities of lensed $z=6$-8
galaxies in our samples with the sizes and masses of stellar systems
in the nearby universe (\S5.3).  Most of the sources have inferred
masses and luminosities that place them in the region of parameter
space where star cluster complexes lie (\S5.2), which occurs midway
between ultra-compact dwarfs and elliptical galaxies.  This suggests
that many low-luminosity galaxies may be dominated by a single star
cluster complex in terms of their observed morphologies.
Nevertheless, we remark that for a small minority of sources in our
sample, their properties are consistent with potentially corresponding
to super star clusters and -- as such -- they could correspond to
proto-globular clusters (Figure~\ref{fig:msre}: see \S5.3).

We combine current constraints on the fraction of especially small
sources behind the HFF clusters with new state-of-the-art constraints
on the $UV$ LF of sources at $z\sim6$ from the HFF clusters (Bouwens
et al.\ 2017b) to derive constraints on the proto-globular cluster LF
at high redshift (\S5.4).  Comparing this LF with predictions from the
recent models from Boylan-Kolchin (2017a, 2017b: but see also Renzini
2017), we find that with current observations from the HFF clusters we
are probably very close to identifying bona-fide globular clusters in
formation in the early universe (if such sources have not been
identified already with our probe or that of Vanzella et al.\ 2017a,
2017b).

For example, the $\xi=1$ model of Boylan-Kolchin (2017a) suggests that
$\sim$3, $\sim$10, and $\sim$3 proto-globular clusters in formation
should be visible in the HFF observations at $-$17.5 mag, $-$16.5 mag,
and $-$15.5 mag.  Tantalizing enough, while our observational results
strongly disfavor objects with proto-globular cluster sizes at
$<$$-$16 mag (see lowest panel of Figure~\ref{fig:gclf}), our results
are plausibly consistent with this model faintward of $-$16 mag, and
in fact this model would suggest that 50\% of the smallest sources in
our $-$15.5 mag selection could correspond to proto-globular clusters.

Despite plausible consistency of our results with the most basic
globular cluster formation models of Boylan-Kolchin (2017a), our
observational results already place strong constraints on more extreme
globular cluster formation scenarios, e.g., with $\xi=10$, ruling out
those scenarios entirely unless the forming star clusters cannot be
picked out amongst the other light in their host galaxy halo (see also
Bastian \& Lardo 2015; Kruijssen 2015; Webb \& Leigh 2015; Martocchia
et al.\ 2017 for further evidence).

As in our previous work (Bouwens et al.\ 2017a) and in other work
identifying especially compact sources in the distant universe
(Vanzella et al.\ 2017a, 2017b), we caution that the present
conclusions depend on the parametric lensing models having predictive
power to magnification factors of $\sim$20 -- which appears quite
likely given the results of e.g. Meneghetti et al.\ 2017, Bouwens et
al.\ 2017b, and Appendix A.

In the future, we plan to extend the present analysis by looking at
the sizes and luminosities of star-forming sources at $z=2$-5 behind
the HFF clusters as well as the $z=6$-8 galaxies behind the final two
HFF clusters when refined public magnification models are available.
Compact star-forming sources identified behind the HFFs represent
compelling compelling targets for spectroscopy with both MUSE and JWST
to gain more insight into the nature of these sources.

\acknowledgements

We acknowledge stimulating discussions with Angela Adamo, Nate
Bastian, Mike Boylan-Kolchin, James Bullock, Rob Crain, Bruce
Elmegreen, Phil Hopkins, Xaiocheng Ma, Mike Norman, Elliot Quataert,
Alvio Renzini, Britton Smith, Eros Vanzella, Dan Weisz, Shelley
Wright, and Tom Zick.  Nate Bastian and Angela Adamo provided us with
extremely valuable feedback on the scientific content and language
used in this paper, especially with regard to star clusters and star
cluster complexes.  We are grateful to Mark Norris for sending us a
compilation of the sizes, luminosities, and masses of many evolved
stellar systems in the nearby universe.  This work utilizes
gravitational lensing models prodcued by PIs Brada{\v c}, Natarajan \&
Kneib (CATS), Merten \& Zitrin, Sharon, and Williams, and the GLAFIC
and Diego groups. This lens modeling was partially funded by the HST
Frontier Fields program conducted by STScI. STScI is operated by the
Association of Universities for Research in Astronomy, Inc. under NASA
contract NAS 5-26555. The lens models were obtained from the Mikulski
Archive for Space Telescopes (MAST).  We acknowledge the support of
NASA grants HST-AR-13252, HST-GO-13872, HST-GO-13792, and NWO grants
600.065.140.11N211 (vrij competitie) and TOP grant TOP1.16.057.

\begin{figure}
\epsscale{0.75}
\plotone{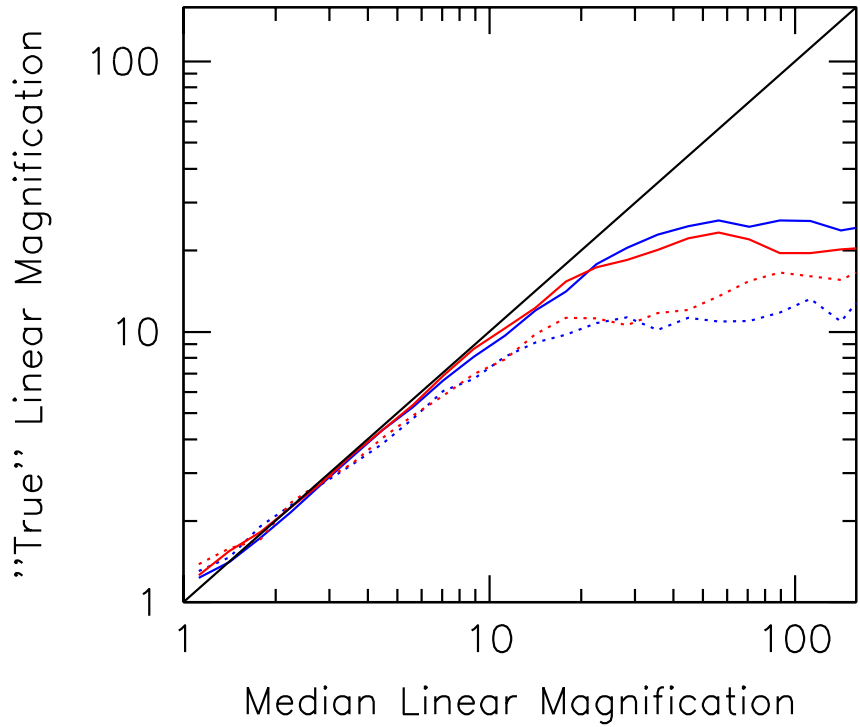}
\caption{An illustration of how well the median linear magnification
  factor likely predicts the actual linear magnification factor.  The
  plotted solid lines show the median of the linear magnification
  factor from the individual parametric models as a function of the
  median linear magnification factor for Abell 2744 (\textit{red}) and
  MACS0416 (\textit{blue}).  The dotted lines show the relationship,
  if the median linear magnification factor from the parametric models
  is compared against the median magnification factor from the
  non-parametric models.  The solid black line is shown for comparison
  to indicate the relationship that would be present for perfect
  predictive power for the lensing models.  The linear magnification
  factors appear to have predictive power to factors of $\sim$20 if we
  assume that the parametric lensing models are taken to represent a
  plausible representation of the actual lensing model and $\sim$10 if
  we assume that the non-parametric models are.  This figure is
  similar in form to Figure 3 from Bouwens et al.\ (2017b), though
  that figure is for the total magnification
  factor.\label{fig:predict}}
\end{figure}

\appendix

\section{A.  Maximum Linear Magnification Factors to Which the Lensing Models Appears to Be Reliable}

While magnification models appear to perform quite well in estimating
the true magnification factors behind lensing clusters (Meneghetti et
al.\ 2017) in the median, these models have difficulty in predicting
the magnification factors very close to the critical curves.  In the
high-magnification regions, the model magnification factors tend to
overpredict the actual magnification factors quite significantly
(e.g., see Figure 3 from Bouwens et al.\ 2017b), e.g., at
$\mu\gtrsim30$.

For the present analysis of sizes, the principal quantity of interest
is not the overall magnification factor, but rather the magnification
along a single spatial dimension.  While the linear magnification
factor was not explicitly considered in the previous analyses of
Meneghetti et al.\ (2017) and Bouwens et al.\ (2017), it should
broadly correlate with the predictive power of the model magnification
factors.

We can quantify the linear magnification factors to which our size
measurements are reliable in the same way we previously determined the
total magnification factors to which our lensing maps were
sufficiently predictive of the total magnification factors (Bouwens et
al.\ 2017b).  As in that work, we alternatively treat one of the
models as if it represented reality and investigated to what extent
the median linear magnification factors from the other models
reproduced the linear magnification factors from the outstanding
model.

We present the results in Figure~\ref{fig:predict} assuming either
parametric models or non-parametric models provided us with the true
magnification and shear maps.  The results in that figure show that
the gravitational lensing models seem capable of predicting the
linear magnification factors $\mu^{1/2} S^{1/2}$ to a value of 20 and
10 assuming that parametric and non-parametric models, respectively,
represented the truth.  Above these values, the median linear
magnification factor no longer strongly correlate with the linear
magnification factors in individual models.

If we assume that the parametric lensing models are plausible
representations of the actual lensing model (as the tests of
Meneghetti et al.\ 2017 suggest), this recommends that we linear
magnification factors to a maximum of 20.  For values above 20, it
suggests we continue to suppose that the actual linear magnification
is 20.

\section{B.  Expected True $UV$ Luminosities vs. Model $UV$ Luminosities}

An important question regards the extent to which the inferred
luminosities of sources behind the HFF clusters actually track their
true luminosities.  Addressing this question is not simple and
requires significant testing through simulation and recovery
experiments.  Previous work included both model-to-model comparisons
(Prieuwe et al.\ 2017; Bouwens et al.\ 2017b) and end-to-end tests
(Meneghetti et al.\ 2017).  These studies have demonstrated that
lensing models appear to be reasonably predictive to a magnification
of 30 in the median, but with 0.4-0.5 dex scatter (see e.g. Figure 3
from Bouwens et al.\ 2017b).  Despite their utility, none of these
tests were framed in terms of the $UV$ luminosity in particular.

\begin{figure}
\epsscale{0.85}
\plotone{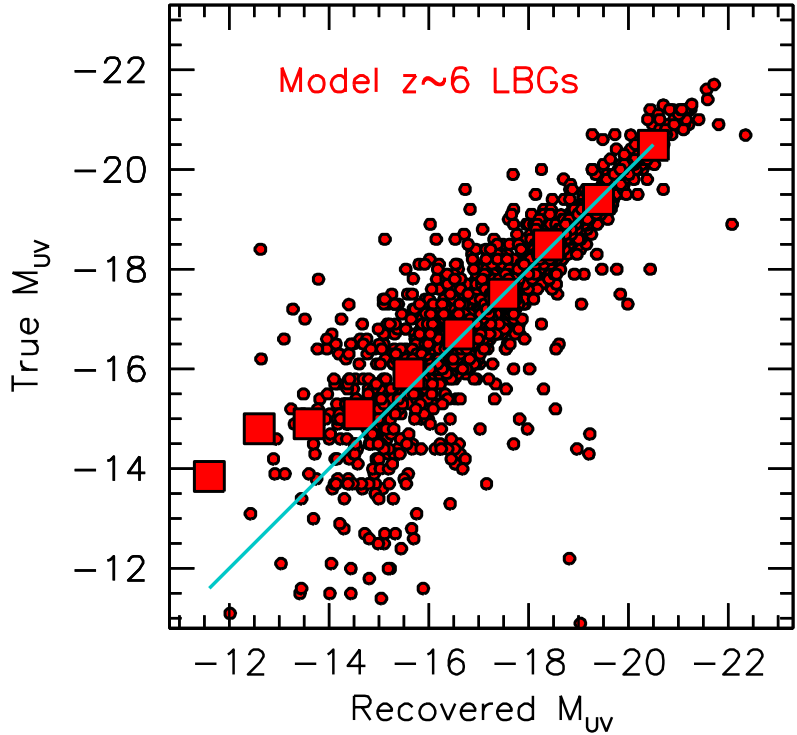}
\caption{``True'' $UV$ luminosity vs. the $UV$ luminosity estimated
  from the median parametric model for sources in a large
  forward-modeling simulation.  The red circles show the original
  model $UV$ luminosities plotted against the recovered $UV$
  luminosities from the median magnification maps.  The red squares
  show the median ``true'' $UV$ luminosity per magnitude bin of
  recovered $UV$ luminosity.  Importantly, the recovered median $UV$
  luminosity is never significantly fainter than $-$15 mag, for
  sources in the $-$14.5 mag, $-$13.5 mag, and $-$12.5 mag
  bins.\label{fig:muvrec}}
\end{figure}

The purpose of this appendix is to look at the extent to which sources
identified as having a given $UV$ luminosity actually have that $UV$
luminosity in the median.  Framing the tests in terms of luminosity
(instead of magnification) is valuable since sources from many
different magnification and apparent magnitude bins contribute to a
given bin in $UV$ luminosity and the total volume within various bins
of $UV$ luminosity varies quite dramaticaly.

To determine how well the inferred $UV$ luminosities actually track
the true $UV$ luminosities, we use the forward-modeling methodology
described in Bouwens et al.\ (2017b).  We use the v4.1 CATS
magnification models to create mock catalogs over each of the first
four HFF clusters, with each source being assigned coordinates and an
apparent magnitude.  Absolute magnitudes are then derived for sources
in these catalogs based on the median magnification from the latest
parametric lensing models.

Both the $UV$ luminosities recovered and the true $UV$ luminosities
are presented in Figure~\ref{fig:muvrec}.  Also shown with the red
squares are how well the luminosities of sources drived from the
median magnification map predict the ``true'' model luminosities.
Interestingly, the luminosities of sources inferred from the median
magnification map track the actual model luminosities brightward of
$-15$ mag, but fail to do so faintward of $-15$ mag.  This suggests
that it may be challenging to quantify with great confidence the
properties or luminosity function of sources fainter than $-$15 mag.

\section{C.  Size measurements for the faintest $z\sim4$ galaxies in the HUDF}

The sizes we infer for many lensed sources in our samples are much
smaller than might be expected based on an extrapolation of the
size-luminosity relation measured from field studies.  While some
trend might have been expected based on surface brightness selection
effects at the low luminosity end of our samples, the fixed surface
density of $z=6$-8 galaxies vs. shear factor (Bouwens et al.\ 2017a)
and high surface density of $z=6$-8 galaxies even in high
magnification regions suggests (\S4.4) that our selections are not
more incomplete to lower luminosity sources than high-luminosity
sources.  This suggests, as we highlighted in our discussions
(\S4.2-4.4), that lower-luminosity star-forming galaxies might
therefore be genuinely small in terms of the observed $UV$ light,
especially when compared with extrapolated size-luminosity relations
from Shibuya et al.\ (2015), for example.

\begin{figure}
\epsscale{0.75}
\plotone{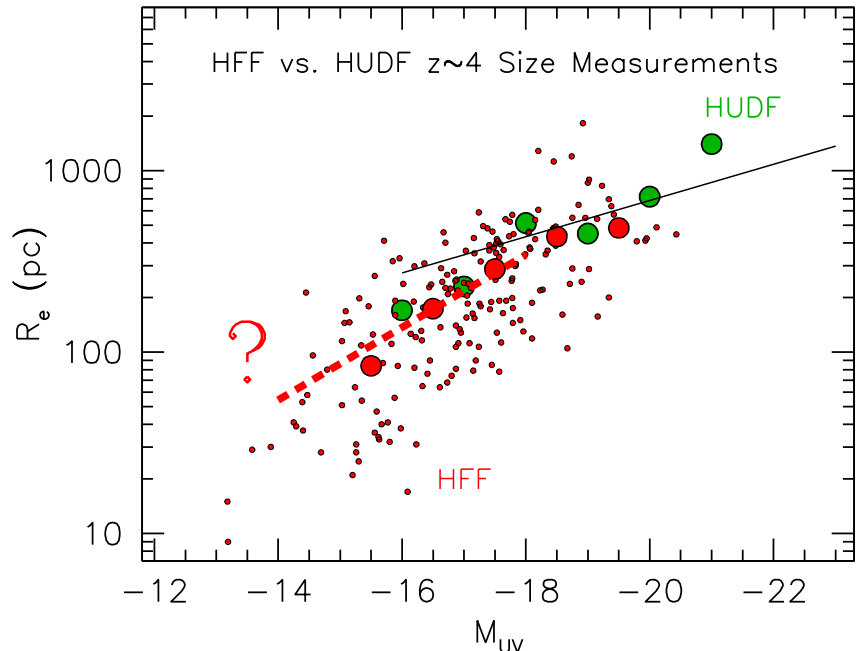}
\caption{A comparison of the median sizes in specific $UV$ luminosity
  bins measured from $z\sim4$ galaxies in the HUDF (\textit{green
    circle}) with the median sizes in specific $UV$ luminosity bins
  measured from the $z\sim4$ sample identified over the MACS0717 and
  MACS1149 HUDF clusters.  The small red points show the size
  measurements for individual sources behind the HFF clusters.  The
  black line shows the Shibuya et al.\ (2015) size-luminosity relation
  at $z\sim4$.  The red dashed line shows one possible size-luminosity
  relation that is both consistent with both our HUDF and HFF
  constraints and shows a steep luminosity dependence (where $r\propto
  L^{0.5}$).  Encouragingly, while the size-luminosity relation at
  $z\sim4$ can only be mapped to moderately low luminosities in the
  HUDF, the median size measurements derived from the HUDF agree with
  our HFF measurements at the faint end of our $z\sim4$ selections.
  Not only does this support the reliability of the size measurements
  made based on our HFF samples are probably reliable, but it suggests
  that the size-luminosity relation may be steeper at the lowest
  luminosities.\label{fig:ok}}
\end{figure}

It is challenging to test this at $z\geq 6$ with current field
samples, as size measurements only probe down to $-18$ mag in our most
sensitive blank field observations the HUDF.  However, at lower
redshifts, i.e., $z\sim4$, one can systematically measure the size of
galaxies to lower luminosities, i.e., $-$16 mag, where one might
expect to see discrepancies between the extrapolated size-luminosity
relation from Shibuya et al.\ (2015) and sources in the HUDF.  

To this end, we made made use of \textsc{galfit} (Peng et al.\ 2002)
to measure source sizes for $z\sim4$ galaxies in HUDF.  Segregating
sources as a function of luminosity, we compute the median sizes as a
function of luminosity.  We present the median sizes in
Figure~\ref{fig:ok} vs. $UV$ luminosity.  For comparison, also shown
in this same figure are the sizes of individual sources from a
$z\sim4$ selection using the HFF clusters
(Bouwens et al.\ 2017, in prep) following the same method as described
in \S4.1.  The median size constraints from our $z\sim4$ HFF selection
are included with the red circles.

Encouragingly enough, the median size measurements we derive from our
$z\sim4$ HUDF sample are in generally good agreement with that
obtained from our $z\sim4$ HFF sample, over the full range of
luminosities.  This suggests that the median size measurements that we
obtain from lensed sources in the HFF data appear to be generally
reliable.  It is also interesting that the median size measurements
for the faintest sources in the HUDF begin to fall below the
size-luminosity relation seen for the brightest galaxies.  While this
may occur due to incompleteness in our selection of star-forming
sources at $z\sim4$ from the HUDF, it could also be providing us with
evidence for a similar break in the size-luminosity relation in our
$z\sim4$ samples.

\section{D.  Size Measurements for Other Sources in Our HFF Samples}

For future comparison studies, we also present in Table~\ref{tab:all}
our inferred size and luminosity measurements for the full set of 307
$z=6$-8 sources utilized in this study, as well as spatial
coordinates, total magnification factors, and linear magnification
factors along the major shear axis.

\begin{deluxetable*}{ccccccc}
\tablecolumns{7}
\tabletypesize{\footnotesize}
\tablecaption{Properties of the Present Compilation $z=6$-8 Sources over the first four HFF clusters\tablenotemark{a}\label{tab:all}}
\tablehead{\colhead{ID} & \colhead{R.A.} & \colhead{Decl} & \colhead{$M_{UV}$} & \colhead{$\mu$} & \colhead{$\mu_{1D}$} & \colhead{$r_e$ (pc)}}
\startdata
A2744I-4205324088 &  00:14:20.54  &  $-$30:24:08.9  & $-$14.6$_{-0.6}^{+0.6}$  & 21.5$_{-8.8}^{+15.3}$  & 6.2$_{-2.5}^{+1.9}$  & 17$_{-13}^{+28}$\\
A2744I-4222023578  &  00:14:22.21  &  $-$30:23:57.9  & $-$15.6$_{-1.2}^{+0.7}$  & 50.1$_{-33.6}^{+41.4}$  & 8.5$_{-3.8}^{+4.4}$  & 31$_{-13}^{+32}$\\
A2744I-4212723104  &  00:14:21.28  &  $-$30:23:10.5  & $-$16.0$_{-0.1}^{+0.1}$  & 6.4$_{-0.8}^{+0.5}$  & 5.7$_{-1.2}^{+0.4}$  & 32$_{-17}^{+24}$\\
A2744Y-4204124034  &  00:14:20.41  &  $-$30:24:03.5  & $-$14.0$_{-1.6}^{+1.2}$  & 75.0$_{-58.0}^{+151.2}$  & 15.1$_{-9.5}^{+24.0}$  & 14$_{-9}^{+29}$\\
M0416I-6055105026  &  04:16:05.52  &  $-$24:05:02.7  & $-$14.6$_{-0.4}^{+1.4}$  & 18.0$_{-6.1}^{+47.0}$  & 16.3$_{-6.1}^{+35.8}$  & 40$_{-27}^{+26}$\\
M0416I-6090604399  &  04:16:09.06  &  $-$24:04:40.0  & $-$15.3$_{-0.5}^{+0.3}$  & 11.6$_{-4.6}^{+3.4}$  & 7.3$_{-3.4}^{+2.4}$  & 32$_{-13}^{+38}$\\
M0416I-6095704260  &  04:16:09.57  &  $-$24:04:26.1  & $-$15.0$_{-0.2}^{+0.3}$  & 12.5$_{-2.5}^{+3.3}$  & 8.0$_{-2.0}^{+2.5}$  & 28$_{-12}^{+19}$\\
M0416I-6120203507  &  04:16:12.02  &  $-$24:03:50.8  & $-$13.6$_{-1.4}^{+1.2}$  & 62.7$_{-45.6}^{+121.6}$  & 57.6$_{-42.7}^{+91.5}$  & 18$_{-11}^{+53}$\\
M0416I-6118103480  &  04:16:11.81  &  $-$24:03:48.1  & $-$15.0$_{-1.0}^{+1.3}$  & 33.6$_{-19.9}^{+81.6}$  & 24.5$_{-15.2}^{+85.1}$  & 16$_{-13}^{+29}$
\enddata
\tablenotetext{a}{Table~\ref{tab:all} is published in its entirety
  in the electronic edition of the Astrophysical Journal.  A portion
  is shown here for guidance regarding its form and content.}
\end{deluxetable*}


\begin{thebibliography}{} 
\bibitem[Adamo et al.(2011)]{2011MNRAS.417.1904A} Adamo, A., {\"O}stlin, G., \& Zackrisson, E.\ 2011, \mnras, 417, 1904
\bibitem[Adamo et al.(2013)]{2013ApJ...766..105A} Adamo, A., {\"O}stlin, G., Bastian, N., et al.\ 2013, \apj, 766, 105 
\bibitem[Alavi et al.(2016)]{2016ApJ...832...56A} Alavi, A., Siana, B., Richard, J., et al.\ 2016, \apj, 832, 56 
\bibitem[Atek et al.(2015)]{2015ApJ...814...69A} Atek, H., Richard, J., Jauzac, M., et al.\ 2015, \apj, 814, 69 
\bibitem[Bastian et al.(2005)]{2005A&A...443...79B} Bastian, N., Gieles, M., Efremov, Y.~N., \& Lamers, H.~J.~G.~L.~M.\ 2005, \aap, 443, 79 
\bibitem[Bastian et al.(2006)]{2006A&A...445..471B} Bastian, N., Emsellem, E., Kissler-Patig, M., \& Maraston, C.\ 2006, \aap, 445, 471 
\bibitem[Bastian(2008)]{2008MNRAS.390..759B} Bastian, N.\ 2008, \mnras, 390, 759 
\bibitem[Bastian et al.(2013)]{2013MNRAS.431.1252B} Bastian, N., Schweizer, F., Goudfrooij, P., Larsen, S.~S., \& Kissler-Patig, M.\ 2013, \mnras, 431, 1252
\bibitem[Bastian \& Lardo(2015)]{2015MNRAS.453..357B} Bastian, N., \& Lardo, C.\ 2015, \mnras, 453, 357 
\bibitem[Beckwith et al.(2006)]{2006AJ....132.1729B} Beckwith, S.~V.~W., 
Stiavelli, M., Koekemoer, A.~M., et al.\ 2006, \aj, 132, 1729 
\bibitem[Bertin and Arnouts (1996)]{1996A&AS..117..393B} Bertin, E.\ and 
Arnouts, S.\ 1996, \aaps, 117, 39
\bibitem[Bouwens et al.(2004)]{2004ApJ...611L...1B} Bouwens, R.~J., 
Illingworth, G.~D., Blakeslee, J.~P., Broadhurst, T.~J., 
\& Franx, M.\ 2004, \apjl, 611, L1
\bibitem[Bouwens et al.(2011)]{2011ApJ...737...90B} Bouwens, R.~J., 
Illingworth, G.~D., Oesch, P.~A., et al.\ 2011, \apj, 737, 90
\bibitem[Bouwens et al.(2014)]{2013arXiv1306.2950B} Bouwens, R.~J., 
Illingworth, G.~D., Oesch, P.~A., et al.\ 2014, \apj, 793, 115
\bibitem[Bouwens et al.(2017)]{2017ApJ...843...41B} Bouwens, R.~J., Illingworth, G.~D., Oesch, P.~A., et al.\ 2017a, \apj, 843, 41 
\bibitem[Bouwens et al.(2017)]{2017ApJ...843..129B} Bouwens, R.~J., Oesch, P.~A., Illingworth, G.~D., Ellis, R.~S., \& Stefanon, M.\ 2017b, \apj, 843, 129
\bibitem[Boylan-Kolchin et al.(2014)]{2014MNRAS.443L..44B} Boylan-Kolchin, M., Bullock, J.~S., \& Garrison-Kimmel, S.\ 2014, \mnras, 443, L44 
\bibitem[Boylan-Kolchin et al.(2015)]{2015MNRAS.453.1503B} Boylan-Kolchin, M., Weisz, D.~R., Johnson, B.~D., et al.\ 2015, \mnras, 453, 1503 
\bibitem[Boylan-Kolchin(2017)]{2017MNRAS.472.3120B} Boylan-Kolchin, M.\ 2017a, \mnras, 472, 3120 
\bibitem[Boylan-Kolchin(2017)]{2017arXiv171100009B} Boylan-Kolchin, M.\ 2017b, \mnras, submitted, arXiv:1711.00009 
\bibitem[Brada{\v c} et al.(2009)]{2009ApJ...706.1201B} Brada{\v c}, M., Treu, T., Applegate, D., et al.\ 2009, \apj, 706, 1201 
\bibitem[Brammer et al.(2008)]{2008ApJ...686.1503B} Brammer, G.~B., van 
Dokkum, P.~G., \& Coppi, P.\ 2008, \apj, 686, 1503
\bibitem[Brodie \& Strader(2006)]{2006ARA&A..44..193B} Brodie, J.~P., \& Strader, J.\ 2006, \araa, 44, 193
\bibitem[Brodie et al.(2011)]{2011AJ....142..199B} Brodie, J.~P., Romanowsky, A.~J., Strader, J., \& Forbes, D.~A.\ 2011, \aj, 142, 199
\bibitem[Cabrera-Ziri et al.(2014)]{2014MNRAS.441.2754C} Cabrera-Ziri, I., Bastian, N., Davies, B., et al.\ 2014, \mnras, 441, 2754 
\bibitem[Cabrera-Ziri et al.(2016)]{2016MNRAS.457..809C} Cabrera-Ziri, I., Bastian, N., Hilker, M., et al.\ 2016, \mnras, 457, 809 
\bibitem[Caminha et al.(2016)]{2016A&A...587A..80C} Caminha, G.~B., Grillo, C., Rosati, P., et al.\ 2016, \aap, 587, A80 
\bibitem[Caminha et al.(2017)]{2017A&A...600A..90C} Caminha, G.~B., Grillo, C., Rosati, P., et al.\ 2017, \aap, 600, A90 
\bibitem[Cappellari et al.(2011)]{2011MNRAS.413..813C} Cappellari, M., Emsellem, E., Krajnovi{\'c}, D., et al.\ 2011, \mnras, 413, 813
\bibitem[Castellano et al.(2016)]{2016ApJ...823L..40C} Castellano, M., Yue, B., Ferrara, A., et al.\ 2016, \apjl, 823, L40 
\bibitem[Coe et al.(2015)]{2015ApJ...800...84C} Coe, D., Bradley, L., \& Zitrin, A.\ 2015, \apj, 800, 84 
\bibitem[Dayal et al.(2014)]{2014MNRAS.445.2545D} Dayal, P., Ferrara, A., Dunlop, J.~S., \& Pacucci, F.\ 2014, \mnras, 445, 2545 
\bibitem[de Jong \& Lacey(2000)]{2000ApJ...545..781D} de Jong, R.~S., \& Lacey, C.\ 2000, \apj, 545, 781 
\bibitem[D'Ercole et al.(2008)]{2008MNRAS.391..825D} D'Ercole, A., Vesperini, E., D'Antona, F., McMillan, S.~L.~W., \& Recchi, S.\ 2008, \mnras, 391, 825 
\bibitem[Dessauges-Zavadsky et al.(2017)]{2017ApJ...836L..22D} Dessauges-Zavadsky, M., Schaerer, D., Cava, A., Mayer, L., \& Tamburello, V.\ 2017, \apjl, 836, L22
\bibitem[Diego et al.(2015)]{2015MNRAS.447.3130D} Diego, J.~M., Broadhurst, T., Molnar, S.~M., Lam, D., \& Lim, J.\ 2015, \mnras, 447, 3130 
\bibitem[Diego et al.(2015)]{2015MNRAS.451.3920D} Diego, J.~M., Broadhurst, T., Zitrin, A., et al.\ 2015, \mnras, 451, 3920 
\bibitem[Diego et al.(2016)]{2016MNRAS.456..356D} Diego, J.~M., Broadhurst, T., Chen, C., et al.\ 2016, \mnras, 456, 356 
\bibitem[Diego et al.(2016)]{2016MNRAS.459.3447D} Diego, J.~M., Broadhurst, T., Wong, J., et al.\ 2016, \mnras, 459, 3447 
\bibitem[Diego et al.(2016)]{2016arXiv160904822D} Diego, J.~M., Schmidt, K.~B., Broadhurst, T., et al.\ 2016, \mnras, submitted, arXiv:1609.04822 
\bibitem[Ellis et al.(2013)]{2013ApJ...763L...7E} Ellis, R.~S., McLure, 
R.~J., Dunlop, J.~S., et al.\ 2013, \apjl, 763, L7 
\bibitem[Elmegreen et al.(2012)]{2012ApJ...757....9E} Elmegreen, B.~G., Malhotra, S., \& Rhoads, J.\ 2012a, \apj, 757, 9 
\bibitem[Elmegreen et al.(2012)]{2012ApJ...747..105E} Elmegreen, B.~G., Zhang, H.-X., \& Hunter, D.~A.\ 2012b, \apj, 747, 105 
\bibitem[Elmegreen et al.(2016)]{2016ApJ...825..145E} Elmegreen, D.~M., Elmegreen, B.~G., S{\'a}nchez Almeida, J., et al.\ 2016, \apj, 825, 145 
\bibitem[Elmegreen(2017)]{2017ApJ...836...80E} Elmegreen, B.~G.\ 2017, \apj, 836, 80 
\bibitem[English \& Freeman(2003)]{2003AJ....125.1124E} English, J., \& Freeman, K.~C.\ 2003, \aj, 125, 1124 
\bibitem[Evstigneeva et al.(2007)]{2007MNRAS.378.1036E} Evstigneeva, E.~A., Drinkwater, M.~J., Jurek, R., et al.\ 2007, \mnras, 378, 1036
\bibitem[Faber(1973)]{1973ApJ...179..423F} Faber, S.~M.\ 1973, \apj, 179, 423 
\bibitem[Ferguson et al.(2004)]{2004ApJ...600L.107F} Ferguson, H.~C., 
Dickinson, M., Giavalisco, M., et al.\ 2004, \apjl, 600, L107
\bibitem[Finlator et al.(2015)]{2015MNRAS.447.2526F} Finlator, K., Thompson, R., Huang, S., et al.\ 2015, \mnras, 447, 2526
\bibitem[Firth et al.(2007)]{2007MNRAS.382.1342F} Firth, P., Drinkwater, M.~J., Evstigneeva, E.~A., et al.\ 2007, \mnras, 382, 1342 
\bibitem[Fisher et al.(2017)]{2017MNRAS.464..491F} Fisher, D.~B., Glazebrook, K., Damjanov, I., et al.\ 2017, \mnras, 464, 491 
\bibitem[Forbes \& Bridges(2010)]{2010MNRAS.404.1203F} Forbes, D.~A., \& Bridges, T.\ 2010, \mnras, 404, 1203 
\bibitem[Franx et al.(1997)]{1997ApJ...486L..75F} Franx, M., Illingworth, G.~D., Kelson, D.~D., van Dokkum, P.~G., \& Tran, K.-V.\ 1997, \apjl, 486, L75
\bibitem[Francis et al.(2012)]{2012MNRAS.425..325F} Francis, K.~J., Drinkwater, M.~J., Chilingarian, I.~V., Bolt, A.~M., \& Firth, P.\ 2012, \mnras, 425, 325
\bibitem[Gnedin(2016)]{2016ApJ...825L..17G} Gnedin, N.~Y.\ 2016, \apjl, 825, L17 
\bibitem[Goddard et al.(2010)]{2010MNRAS.405..857G} Goddard, Q.~E., Bastian, N., \& Kennicutt, R.~C.\ 2010, \mnras, 405, 857
\bibitem[Graus et al.(2016)]{2016MNRAS.456..477G} Graus, A.~S., Bullock, J.~S., Boylan-Kolchin, M., \& Weisz, D.~R.\ 2016, \mnras, 456, 477
\bibitem[Grazian et 
al.(2011)]{2011A&A...532A..33G} Grazian, A., Castellano, M., Koekemoer, A.~M., et al.\ 2011, \aap, 532, A33 
\bibitem[Grazian et 
al.(2012)]{2012A&A...547A..51G} Grazian, A., Castellano, M., Fontana, A., et al.\ 2012, \aap, 547, A51 
\bibitem[Harris et al.(2013)]{2013ApJ...772...82H} Harris, W.~E., Harris, G.~L.~H., \& Alessi, M.\ 2013, \apj, 772, 82 
\bibitem[Ha{\c s}egan et al.(2005)]{2005ApJ...627..203H} Ha{\c s}egan, M., Jord{\'a}n, A., C{\^o}t{\'e}, P., et al.\ 2005, \apj, 627, 203
\bibitem[Holwerda et al.(2015)]{2015ApJ...808....6H} Holwerda, B.~W., Bouwens, R., Oesch, P., et al.\ 2015, \apj, 808, 6 
\bibitem[Huang et al.(2013)]{2013ApJ...765...68H} Huang, K.-H., Ferguson, H.~C., Ravindranath, S., \& Su, J.\ 2013, \apj, 765, 68 
\bibitem[Illingworth et al.(2013)]{2013ApJS..209....6I} Illingworth, G.~D., 
Magee, D., Oesch, P.~A., et al.\ 2013, \apjs, 209, 6
\bibitem[Ishigaki et al.(2015)]{2015ApJ...799...12I} Ishigaki, M., Kawamata, R., Ouchi, M., et al.\ 2015, \apj, 799, 12 
\bibitem[Ishigaki et al.(2017)]{2017arXiv170204867I} Ishigaki, M., Kawamata, R., Ouchi, M., Oguri, M., \& Shimasaku, K.\ 2017, \apj, submitted, arXiv:1702.04867 
\bibitem[Jauzac et al.(2015)]{2015MNRAS.446.4132J} Jauzac, M., Jullo, E., Eckert, D., et al.\ 2015a, \mnras, 446, 4132 
\bibitem[Jauzac et al.(2015)]{2015MNRAS.452.1437J} Jauzac, M., Richard, J., Jullo, E., et al.\ 2015b, \mnras, 452, 1437 
\bibitem[Jauzac et al.(2016)]{2016MNRAS.457.2029J} Jauzac, M., Richard, J., Limousin, M., et al.\ 2016, \mnras, 457, 2029 
\bibitem[Johnson et al.(2014)]{2014ApJ...797...48J} Johnson, T.~L., Sharon, K., Bayliss, M.~B., et al.\ 2014, \apj, 797, 48 
\bibitem[Johnson et al.(2017)]{2017arXiv170700706J} Johnson, T.~L., Rigby, J.~R., Sharon, K., et al.\ 2017, arXiv:1707.00706 
\bibitem[Jones et al.(2010)]{2010MNRAS.404.1247J} Jones, T.~A., Swinbank, A.~M., Ellis, R.~S., Richard, J., \& Stark, D.~P.\ 2010, \mnras, 404, 1247 
\bibitem[Jullo \& Kneib(2009)]{2009MNRAS.395.1319J} Jullo, E., \& Kneib, J.-P.\ 2009, \mnras, 395, 1319 
\bibitem[Lada \& Lada(2003)]{2003ARA&A..41...57L} Lada, C.~J., \& Lada, E.~A.\ 2003, \araa, 41, 57 
\bibitem[Lagattuta et al.(2017)]{2017MNRAS.469.3946L} Lagattuta, D.~J., Richard, J., Cl{\'e}ment, B., et al.\ 2017, \mnras, 469, 3946
\bibitem[Lam et al.(2014)]{2014ApJ...797...98L} Lam, D., Broadhurst, T., Diego, J.~M., et al.\ 2014, \apj, 797, 98 
\bibitem[Laporte et al.(2016)]{2016ApJ...820...98L} Laporte, N., Infante, L., Troncoso Iribarren, P., et al.\ 2016, \apj, 820, 98 
\bibitem[Liesenborgs et al.(2006)]{2006MNRAS.367.1209L} Liesenborgs, J., De Rijcke, S., \& Dejonghe, H.\ 2006, \mnras, 367, 1209 
\bibitem[Limousin et al.(2016)]{2016A&A...588A..99L} Limousin, M., Richard, J., Jullo, E., et al.\ 2016, \aap, 588, A99 
\bibitem[Liu et al.(2017)]{2017MNRAS.465.3134L} Liu, C., Mutch, S.~J., Poole, G.~B., et al.\ 2017, \mnras, 465, 3134 
\bibitem[Liu et al.(2016)]{2016MNRAS.462..235L} Liu, C., Mutch, S.~J., Angel, P.~W., et al.\ 2016, \mnras, 462, 235
\bibitem[Livermore et al.(2012)]{2012MNRAS.427..688L} Livermore, R.~C., Jones, T., Richard, J., et al.\ 2012, \mnras, 427, 688
\bibitem[Livermore et al.(2015)]{2015MNRAS.450.1812L} Livermore, R.~C., Jones, T.~A., Richard, J., et al.\ 2015, \mnras, 450, 1812
\bibitem[Livermore et al. (2017)]{livermore} Livermore, R., Finkelstein, S., Lotz,
  J. 2017, ApJ, 835, 113
\bibitem[Lotz et al.(2017)]{2017ApJ...837...97L} Lotz, J.~M., Koekemoer, A., Coe, D., et al.\ 2017, \apj, 837, 97 
\bibitem[Kawamata et al.(2015)]{2015ApJ...804..103K} Kawamata, R., Ishigaki, M., Shimasaku, K., Oguri, M., \& Ouchi, M.\ 2015, \apj, 804, 103 
\bibitem[Kawamata et al.(2016)]{2016ApJ...819..114K} Kawamata, R., Oguri, M., Ishigaki, M., Shimasaku, K., \& Ouchi, M.\ 2016, \apj, 819, 114 
\bibitem[Kawamata et al.(2017)]{2017arXiv171007301K} Kawamata, R., Ishigaki, M., Shimasaku, K., et al.\ 2017, \apj, submitted, arXiv:1710.07301 
\bibitem[Keeton(2010)]{2010GReGr..42.2151K} Keeton, C.~R.\ 2010, General Relativity and Gravitation, 42, 2151 
\bibitem[Kennicutt et al.(2003)]{2003PASP..115..928K} Kennicutt, R.~C., Jr., Armus, L., Bendo, G., et al.\ 2003, \pasp, 115, 928
\bibitem[Kravtsov(2013)]{2013ApJ...764L..31K} Kravtsov, A.~V.\ 2013, \apjl, 764, L31 
\bibitem[Kruijssen(2014)]{2014CQGra..31x4006K} Kruijssen, J.~M.~D.\ 2014, Classical and Quantum Gravity, 31, 244006 
\bibitem[Kruijssen(2015)]{2015MNRAS.454.1658K} Kruijssen, J.~M.~D.\ 2015, \mnras, 454, 1658 
\bibitem[Ma et al.(2017)]{2017arXiv171000008M} Ma, X., Hopkins, P.~F., Boylan-Kolchin, M., et al.\ 2017, \mnras, \submitted, arXiv:1710.00008 
\bibitem[Mahler et al.(2017)]{2017arXiv170206962M} Mahler, G., Richard, J., Cl{\'e}ment, B., et al.\ 2017, \mnras, submitted, arXiv:1702.06962 
\bibitem[Maraston et al.(2004)]{2004A&A...416..467M} Maraston, C., Bastian, N., Saglia, R.~P., et al.\ 2004, \aap, 416, 467 
\bibitem[Martocchia et al.(2017)]{2017MNRAS.468.3150M} Martocchia, S., Bastian, N., Usher, C., et al.\ 2017, \mnras, 468, 3150
\bibitem[McDermid et al.(2015)]{2015MNRAS.448.3484M} McDermid, R.~M., Alatalo, K., Blitz, L., et al.\ 2015, \mnras, 448, 3484
\bibitem[McLure et al.(2013)]{2013MNRAS.432.2696M} McLure, R.~J., Dunlop, 
J.~S., Bowler, R.~A.~A., et al.\ 2013, \mnras, 432, 2696
\bibitem[Meneghetti et al.(2017)]{2017MNRAS.472.3177M} Meneghetti, M., Natarajan, P., Coe, D., et al.\ 2017, \mnras, 472, 3177 
\bibitem[Merlin et al.(2016)]{2016A&A...590A..30M} Merlin, E., Amor{\'{\i}}n, R., Castellano, M., et al.\ 2016, \aap, 590, A30 
\bibitem[Meurer et al.(1995)]{1995AJ....110.2665M} Meurer, G.~R., Heckman, T.~M., Leitherer, C., et al.\ 1995, \aj, 110, 2665 
\bibitem[Mieske et al.(2007)]{2007A&A...472..111M} Mieske, S., Hilker, M., Jord{\'a}n, A., Infante, L., \& Kissler-Patig, M.\ 2007, \aap, 472, 111
\bibitem[Misgeld et al.(2008)]{2008A&A...486..697M} Misgeld, I., Mieske, S., \& Hilker, M.\ 2008, \aap, 486, 697 
\bibitem[Misgeld et al.(2011)]{2011A&A...531A...4M} Misgeld, I., Mieske, S., Hilker, M., et al.\ 2011, \aap, 531, A4
\bibitem[Mosleh et al.(2012)]{2012ApJ...756L..12M} Mosleh, M., Williams, R.~J., Franx, M., et al.\ 2012, \apjl, 756, L12 
\bibitem[Murray(2009)]{2009ApJ...691..946M} Murray, N.\ 2009, \apj, 691, 946 
\bibitem[Norris et al.(2014)]{2014norris} Norris, M., Kannappan, S., Forbes, D.A., et al.\ 2014, \mnras, arXiv:1406.6065v1
\bibitem[Ocvirk et al.(2016)]{2016MNRAS.463.1462O} Ocvirk, P., Gillet, N., Shapiro, P.~R., et al.\ 2016, \mnras, 463, 1462 
\bibitem[Oesch et al.(2010)]{2010ApJ...709L..16O} Oesch, P.~A., Bouwens, 
R.~J., Illingworth, G.~D., et al.\ 2010a, \apjl, 709, L16
\bibitem[Oesch et al.(2015)]{2015ApJ...808..104O} Oesch, P.~A., Bouwens, R.~J., Illingworth, G.~D., et al.\ 2015, \apj, 808, 104
\bibitem[Oguri(2010)]{2010PASJ...62.1017O} Oguri, M.\ 2010, \pasj, 62, 1017 
\bibitem[Oke \& Gunn(1983)]{1983ApJ...266..713O} Oke, J.~B., \& Gunn, 
J.~E.\ 1983, \apj, 266, 713 
\bibitem[Ono et al.(2013)]{2013ApJ...777..155O} Ono, Y., Ouchi, M., 
Curtis-Lake, E., et al.\ 2013, \apj, 777, 155
\bibitem[Owers et al.(2011)]{2011ApJ...728...27O} Owers, M.~S., Randall, S.~W., Nulsen, P.~E.~J., et al.\ 2011, \apj, 728, 27 
\bibitem[Peng et al.(2002)]{2002AJ....124..266P} Peng, C.~Y., Ho, L.~C., 
Impey, C.~D., \& Rix, H.-W.\ 2002, \aj, 124, 266
\bibitem[Planck Collaboration et al.(2015)]{2015arXiv150201589P} Planck 
Collaboration, Ade, P.~A.~R., Aghanim, N., et al.\ 2015, arXiv:1502.01589 [PC15]
\bibitem[Postman et al.(2012)]{2012ApJS..199...25P} Postman, M., Coe, D., 
Ben{\'{\i}}tez, N., et al.\ 2012, \apjs, 199, 25
\bibitem[Priewe et al.(2016)]{2016arXiv160507621P} Priewe, J., Williams, L.~L.~R., Liesenborgs, J., Coe, D., \& Rodney, S.~A.\ 2016, \mnras, submitted, arXiv:1605.07621 
\bibitem[Rejkuba et al.(2007)]{2007A&A...469..147R} Rejkuba, M., Dubath, P., Minniti, D., \& Meylan, G.\ 2007, \aap, 469, 147 
\bibitem[Renzini et al.(2015)]{2015MNRAS.454.4197R} Renzini, A., D'Antona, F., Cassisi, S., et al.\ 2015, \mnras, 454, 4197 
\bibitem[Renzini(2017)]{2017MNRAS.469L..63R} Renzini, A.\ 2017, \mnras, 469, L63 
\bibitem[Rodr{\'{\i}}guez-Zaur{\'{\i}}n et al.(2011)]{2011A&A...527A..60R} Rodr{\'{\i}}guez-Zaur{\'{\i}}n, J., Arribas, S., Monreal-Ibero, A., et al.\ 2011, \aap, 527, A60 
\bibitem[Ribeiro et al.(2016)]{2016arXiv161105869R} Ribeiro, B., Le F{\`e}vre, O., Cassata, P., et al.\ 2016, \aap, submitted, arXiv:1611.05869
\bibitem[Richard et al.(2014)]{2014MNRAS.444..268R} Richard, J., Jauzac, M., Limousin, M., et al.\ 2014, \mnras, 444, 268 
\bibitem[Schaerer \& Charbonnel(2011)]{2011MNRAS.413.2297S} Schaerer, D., \& Charbonnel, C.\ 2011, \mnras, 413, 2297 

\bibitem[Schmidt et al.(2014)]{2014ApJ...786...57S} Schmidt, K.~B., Treu, 
T., Trenti, M., et al.\ 2014, \apj, 786, 57 
\bibitem[Sebesta et al.(2016)]{2015arXiv150708960S} Sebesta, K., Williams, L.~L.~R., Mohammed, I., Saha, P., \& Liesenborgs, J.\ 2016, \mnras, 461, 2126 
\bibitem[Shibuya et al.(2015)]{2015ApJS..219...15S} Shibuya, T., Ouchi, M., \& Harikane, Y.\ 2015, \apjs, 219, 15 
\bibitem[Silva-Villa \& Larsen(2011)]{2011A&A...529A..25S} Silva-Villa, E., \& Larsen, S.~S.\ 2011, \aap, 529, A25
\bibitem[Simon \& Geha(2007)]{2007ApJ...670..313S} Simon, J.~D., \& Geha, M.\ 2007, \apj, 670, 313
\bibitem[Spitler \& Forbes(2009)]{2009MNRAS.392L...1S} Spitler, L.~R., \& Forbes, D.~A.\ 2009, \mnras, 392, L1 

\bibitem[Swinbank et al.(2009)]{2009MNRAS.400.1121S} Swinbank, A.~M., Webb, T.~M., Richard, J., et al.\ 2009, \mnras, 400, 1121
\bibitem[Swinbank et al.(2012)]{2012ApJ...760..130S} Swinbank, A.~M., Smail, I., Sobral, D., et al.\ 2012, \apj, 760, 130 
\bibitem[Taghizadeh-Popp et al.(2015)]{2015ApJ...801...14T} Taghizadeh-Popp, M., Fall, S.~M., White, R.~L., \& Szalay, A.~S.\ 2015, \apj, 801, 14 
\bibitem[van der Wel et al.(2014)]{2014ApJ...788...28V} van der Wel, A., Franx, M., van Dokkum, P.~G., et al.\ 2014, \apj, 788, 28 
\bibitem[van Dokkum et al.(2015)]{2015ApJ...798L..45V} van Dokkum, P.~G., Abraham, R., Merritt, A., et al.\ 2015, \apjl, 798, L45
\bibitem[Vanzella et al.(2014)]{2014ApJ...783L..12V} Vanzella, E., Fontana, A., Zitrin, A., et al.\ 2014, \apjl, 783, L12
\bibitem[Vanzella et al.(2017)]{2017MNRAS.467.4304V} Vanzella, E., Calura, F., Meneghetti, M., et al.\ 2017a, \mnras, 467, 4304
\bibitem[Vanzella et al.(2017)]{2017ApJ...842...47V} Vanzella, E., Castellano, M., Meneghetti, M., et al.\ 2017b, \apj, 842, 47
\bibitem[Vanzi et al.(2008)]{2008A&A...486..393V} Vanzi, L., Cresci, G., Telles, E., \& Melnick, J.\ 2008, \aap, 486, 393 
\bibitem[Webb \& Leigh(2015)]{2015MNRAS.453.3278W} Webb, J.~J., \& Leigh, N.~W.~C.\ 2015, \mnras, 453, 3278 
\bibitem[Weisz et al.(2014)]{2014ApJ...794L...3W} Weisz, D.~R., Johnson, B.~D., \& Conroy, C.\ 2014, \apjl, 794, L3 
\bibitem[Wisnioski et al.(2012)]{2012MNRAS.422.3339W} Wisnioski, E., Glazebrook, K., Blake, C., et al.\ 2012, \mnras, 422, 3339
\bibitem[Yue et al.(2016)]{2016arXiv160401314Y} Yue, B., Ferrara, A., \& Xu, Y.\ 2016, arXiv:1604.01314 
\bibitem[Zitrin et al.(2011)]{2011MNRAS.413.1753Z} Zitrin, A., Broadhurst, T., Coe, D., et al.\ 2011, \mnras, 413, 1753 
\bibitem[Zitrin et al.(2012)]{2012MNRAS.423.2308Z} Zitrin, A., Broadhurst, T., Bartelmann, M., et al.\ 2012, \mnras, 423, 2308 
\bibitem[Zitrin et al.(2013)]{2013ApJ...762L..30Z} Zitrin, A., Meneghetti, M., Umetsu, K., et al.\ 2013, \apjl, 762, L30 
\bibitem[Zitrin et al.(2015)]{2015ApJ...801...44Z} Zitrin, A., Fabris, A., Merten, J., et al.\ 2015, \apj, 801, 44 
\end{thebibliography}
\end{document}